\documentclass[fleqn]{llncs}
\usepackage{amssymb}
\usepackage{color}
\usepackage{boxedminipage}

\begin{document}

\title{Formal Verification of `Programming to Interfaces' Programs
\thanks{This paper is supported by the Chinese National 863 Project, NO.2011AA010103}}
\author{ZHAO Jianhua\thanks{The corresponding author.}, LI Xuandong}

\institute{State Key Laboratory of Novel Software Technology\\
    Dept. of Computer Sci. and Tech. Nanjing University\\
    Nanjing, Jiangsu, P.R.China 210093\\
    \{zhaojh,lxd\}@nju.edu.cn}
\maketitle

\newcommand{\seml}{[\![}
\newcommand{\semr}{]\!]}
\newcommand{\fldacc}{\mbox{$\rightarrow$}}
\newcommand{\outlying}{\texttt{Outlying}}
\begin{abstract}
This paper presents a formal approach to specify and verify object-oriented programs written in the
`programming to interfaces' paradigm. Besides the methods to be invoked by its clients, an interface
also declares a set of abstract function/predicate symbols, together with a set of constraints
on these symbols. For each method declared in this interface, a specification template is given using these abstract symbols.
A class implementing this interface can give its own definitions to the abstract symbols, as long as all the constraints are satisfied.
This class implements all the methods declared in the interface such that
the method specification templates declared in the interface are satisfied w.r.t. the definitions of the abstract function symbols in this class.
Based on the constraints on the abstract symbols, the client code using interfaces can be specified and verified precisely without knowing
what classes implement these interfaces. Given more information about the implementing classes, the
specifications of the client code can be specialized into more precise ones without re-verifying the client code.

Several commonly used interfaces and their implementations (including Iterator, Observer, Comparable, and Comparator) are used to
demonstrate that the approach in this paper is both precise and flexible.
\end{abstract}

\section{Introduction}
One of the important programming paradigms of OO programming is `programming to interfaces'.
Programmers can use an interface without knowing the details of its implementations.
This programming paradigm decouples the code using the interfaces and the implementations of these
interfaces. It also makes programs more flexible, because programmers can make client code using the interface fulfill different functional
features using different implementations of an interface, without modifying the client code.
So this paradigm is widely used in software developments, and supported by many modern
OO programming languages. In Java, \texttt{interface} is an important concept used to support this paradigm.
In C++, abstract base classes play the similar role as \texttt{interface} does in Java.
There are already a number of research works on how to deal with inheritance and method overriding.
However, these techniques are not powerful enough to take the full advantage of the `programming to interfaces' paradigm.

Now we take the interface \texttt{java.lang.Comparable} in the standard package of Java as an example to show
why these formal techniques are not powerful enough. The interface \texttt{Comparable}
has one method \texttt{CompareTo}, which compares an object with the \textbf{this} object.
A class implementing this interface must define this method. According to the standard documentation of Java, the implementor must ensure that the following
formulas hold for any $x,y,z$.
\begin{itemize}
\item  $\texttt{sgn}(x\fldacc\texttt{compareTo}(y)) = -\texttt{sgn}(y\fldacc\texttt{compareTo}(x))$;
\item  $(x\fldacc\texttt{compareTo}(y)>0 \land y\fldacc\texttt{compareTo}(z)>0)\Rightarrow x\fldacc\texttt{compareTo}(z)>0$;
\item  $(x\fldacc\texttt{compareTo}(y)= 0)\Rightarrow (\texttt{sgn}(x\fldacc\texttt{compareTo}(z)) = \texttt{sgn}(y\fldacc\texttt{compareTo}(z)))$.
\end{itemize}
Here \texttt{sgn} is a function which yields -1, 0 and +1 respectively when the parameter is less than, equal to or greater than 0.

The above requirements mean that the method \texttt{compareTo} induces a total order over the objects. The following
code returns the `smaller' object of the two parameters, given that the two parameters are of the same class implementing the interface \texttt{Comparable}.
\begin{center}
\begin{minipage}{0.6\textwidth}
\scriptsize
\texttt{Comparable} TheSmallerOne(\texttt{Comparable} o1; \texttt{Comparable} o2)\\
\{\\
\mbox{}\ \ \ \ \textbf{if} (o1\fldacc\texttt{compareTo}(o2)$>$0)\ \ \textbf{return} o2;\ \ \textbf{else}\ \ \textbf{return} o1;\ \ \\
\}
\end{minipage}
\end{center}
Different implementations of the method \texttt{compareTo} give different meanings to the word `smaller'.

To support the `programming to interfaces' paradigm, the client code using interfaces must be specified and verified under the open-world assumption, i.e. without knowing what classes implement the interfaces. This is critical because
programmers should be able to add a new implementation to an interface without having to re-verify the client code using this interface. Further more, the specification of the client code must be precise and flexible enough, such that programmers can conclude that their implementation of the interface
makes the client code fulfill the functional features as expected.
For example, suppose that a class \texttt{Point} implements the interface \texttt{Comparable}. A \texttt{Point} object represents a point in the X-Y plate.  The method \texttt{compareTo} defined in \texttt{Point} compares two points by their distances to the original point. The programmer should be able to conclude that the above method returns the point closer to the original point, if the method is invoked using two points as the real parameters.

In the programming language C++, interfaces are written as abstract classes, and they are implemented by
their subclasses. People can use the \texttt{inheritance} facility to give different implementations to an abstract class. To avoid re-verification of the client code of the abstract classes, researchers deal with \texttt{inheritance} based on the Liskov Substitution Principle (LSP) sub-typing rule\cite{Liskov:1994:BNS:197320.197383}. For a method overridden in a subclass, the overriding method should have a behavioral sub-typing relation with the overridden one. Roughly speaking, the overriding method should have a weaker precondition and a stronger post-condition. However, this approach can not be applied in the `programming to interfaces' paradigm because of two reasons.
\begin{itemize}
\item An interface (or an abstract class) declares no member variable. So people can not specify how the member variables are accessed and affected by a method when an interface is designed. Specifying a method based only on the parameters and return values is not precise enough. For example, the return value of \texttt{compareTo} depends on the member variables of both the \textbf{this} object and the real parameter.
\item The behavioral sub-typing is too restrictive for the `programming to interfaces' paradigm. In many cases, different implementations of an interface SHOULD behave differently such that the client code using the interface may have different functional features as expected. So even if precise specifications of the interface methods are given somehow, the behavioral sub-typing relation may hurt the flexibility of the paradigm. For example, different implementations of the method compareTo should make the method \texttt{TheSmallOne} choose the `smaller' object differently.
\end{itemize}

In this paper, an approach is presented to specify and verify programs written in the `programming to interfaces' paradigm. The main contributions of this paper are as follows.
\begin{itemize}
\item \textbf{Abstract but precise specifications of interface methods.} The interface methods are specified through a set of function/predicate symbols and their constraints declared in the interface. These symbols are abstract in the interface, and to be defined in the classes implementing this interface. Because of the constraints, the specification templates given in an interface is precise enough to reason about the client code using this interface.
\item \textbf{A flexible implementation relation between interfaces and classes.} To implement an interface, a class must define all the function symbols and implement all  the methods declared in the interface.  The only two restrictions are (1) the function/predicate symbol definitions must satisfy the constraints declared in the interface; (2) the method implementations should satisfy the specification templates declared in the interface w.r.t. the symbol definitions in this class.
    When the idea in this paper is applied to deal with class inheritance, the LSP sub-typing rule is a special case of our rule.
\item \textbf{Verifications of the client code under the open-world assumption.} No knowledge about the implementing classes is needed during the verification of the client
    code of interfaces. Furthermore, when more information about the runtime classes is given, the specifications of the client code can be specialized to more precise ones.
\end{itemize}

The rest part of this paper is organized as follows. Section~\ref{SEC-BRIEF-INTRODUCTION-LANGUAGE} gives a brief introduction on the small language used in this paper. The syntax of interface is described in Section~\ref{SEC-INTERFACE}. The syntax of class definitions is given in  Section~\ref{SEC-CLASS}. This section also
discusses how to verify the implementation relation between a class and an interface, especially how to verify
a method w.r.t. its specifications. The types and expressions associated with interfaces and classes are discussed in Section~\ref{SEC-TYPE-EXPRESSION}. The scope memory rules for these expressions are also given in this section.
In Section~\ref{SEC-STATEMENT}, the syntax and proof rules of the statements associated with classes and interfaces are given. The approach to verify
a client code using interfaces is presented in Section~\ref{SEC-OPEN-WORLD}. Section~\ref{SEC-DISCUSSION} discusses briefly how to deal with inheritance using the approach
presented in this paper. Section~\ref{SEC-CONCLUSION} compares the method in this paper with related works and concludes this paper.
In Appendixes, we give more examples of some widely-used interfaces, their implementations,
and the client codes using these interfaces.

\section{A brief introduction to the small language}\label{SEC-BRIEF-INTRODUCTION-LANGUAGE}
This section gives a brief description about the small OO language used in this paper. More details about this language will be given later.

A program of the small language consists of a set of interface declarations and class definitions.
An interface can be implemented by one or more classes, while a class can implement zero or more interfaces.
In these interface declarations and class definitions, code are given together with their specifications.

An interface declares a set of methods that can be invoked by its users and a set of function (predicate) symbols used in method specifications.
All these methods and function symbols are polymorphic, i.e. the classes implementing the interface can given their own definitions to them.
For a method invocation $e\fldacc m()$, the method definition given in the runtime class of $e$ is invoked. Similarly, for a term $e\fldacc f()$ in specifications,
where $f$ is a function symbol declared in an interface, $f$ also refers to the definition of $f$ in the runtime class of $e$.

For these polymorphic symbols, class-prefixes are used to specify which definitions they refer to. For example, $C\mbox{::}f$ refers to
the definition of $f$ in the class $C$. We use $\textbf{classOf}(e)$ to denote the dynamic class of $e$, and $\textbf{classOf}(e)\mbox{::}f$ refers to the
definition of $f$ in the dynamic class of $e$.
In interface declarations, we use the keyword \textbf{theClass} to represent the class implementing this interface. Intuitively speaking,
a sub-expression $\textbf{theClass}\mbox{::}f(\overline y)$ in the declaration of an interface $I$ has its counter
part $C\mbox{::}f(\overline y)$ in the definition of a class $C$ implementing $I$.

For each method $m$ declared in an interface $I$, the templates of the precondition and the post-condition of $m$ are given in $I$.
A template is a formula containing the keyword \textbf{theClass}. For a class $C$ implementing $I$, the specification of $m$ is derived
by substituting $\textbf{theClass}$  with $C$ in the templates.

An interface also declares a set of constraint templates about the function symbols declared in the interface.
In a class $C$ implementing $I$, a set of constraints are derived by substituting \textbf{theClass} with $C$ in these constraint templates.
The function symbol definitions in $C$ must satisfy these constraints.

A class definition $C$ defines a set of methods and function symbols. For each interface $I$ implemented by the class $C$, all the methods and function
symbols declared in $I$ should be defined in $C$.

To verify a program, we have the following two kinds of proof obligations.
\begin{itemize}
\item A class $C$ defines all the methods and function symbols declared in each interface implemented by $C$. Further more, the function symbol definitions must satisfy all the  constraints derived by substituting \textbf{theClass} with $C$ in the constraint templates in the interfaces.
\item Each method defined in a class $C$ should satisfy its specifications. For a method $m$ declared in an interface $I$ implemented by $C$, its specification in $C$ is derived by substituting \textbf{theClass} with $C$ in the specification template of $m$ in $I$.

\end{itemize}

\section{The interface declarations}\label{SEC-INTERFACE}
An interface declares a set of methods that can be invoked by its users. The templates of the pre-conditions and post-conditions of these methods are also given in the interface declaration.
An interface also declares a set of function symbols which are
used to specify the methods of this interface.
A set of constraint templates about the function symbols are also given in the interface.
The templates for pre-/post-conditions and constraints are formulas containing the keyword \textbf{theClass}. It will be substituted with the concrete class name in the
class definitions implementing this interface.


\begin{figure}
\begin{center}
\begin{minipage}{0.8\textwidth}
ISpec ::= iName `\{' fSymDecs Constraints  methodDecs `\}'\\
fSymDecs ::= $\epsilon$ $|$ \textbf{funcs} `:' fDec (';' fDec)$^+$\\
fDec  ::= ( \textbf{static} $|$ \textbf{attrib} $|$ $\epsilon$ ) Type fName `(' fPara `)'\\
Constraints  ::= $\epsilon$ $|$ \textbf{cons} `:' Formula ( `;' Formula)$^*$\\
methodDecs ::= $\epsilon$ $|$ \textbf{methods} `:' (MSpec)$^+$\\
MSpec  ::= Type fname `(' fPara `)' \textbf{pre} Formula \textbf{post} Formula `;'
\end{minipage}
\end{center}
\caption{The grammar of interface specifications}\label{SYN-INTERFACE-SPECIFICATION}
\end{figure}

The grammar of interface specifications is depicted in Fig.~\ref{SYN-INTERFACE-SPECIFICATION}.
The meta expression $(\dots)^*$  means that the component in the brackets can repeat for zero or many times.
The meta expression $(\dots)^+$  means that the component in the brackets can repeat for one or many times. An interface declaration
consists of three parts: the function symbol declarations, the constraint templates about these function symbols, and the method declarations and specifications.
In the grammar, `iName' and `fName' are identifiers, representing interface names and function symbol names respectively.
`Type' represents type-expressions, `fPara' represents formal parameter declarations. `Formula' represents boolean-typed expressions.
The syntax of Type and Formula are given in \cite{DBLP:conf/ictac/ZhaoL13}.

\begin{example}
An interface \texttt{Comparable} is declared in Fig.~\ref{FIG-COMPARABLE}.
This interface is similar to the one given in the standard Java library java.lang.
It declares two function symbols \texttt{VALUE} and \texttt{LE}. Three constraint templates about \texttt{VALUE} and \texttt{LE} are given in this interface.
This interface declares one method \texttt{compareTo}. The pre-/post-condition templates of \texttt{compareTo} are given
using the function symbols \texttt{VALUE} and \texttt{LE}.


%
\hfill$\square$
\end{example}

\begin{figure}
\begin{center}
\begin{boxedminipage}{1\textwidth}
\scriptsize
\begin{tabbing}
\ \ \ \ \=\ \ \ \ \=\ \ \ \ \=\ \ \ \=\\
\textbf{interface} \texttt{Comparable}\\
\{\\
\textbf{funcs:}\\
    \>\textbf{attrib} \textbf{int}  \texttt{VALUE}();\\
    \>\textbf{static} \textbf{bool} \texttt{LE}(\textbf{int} v1, \textbf{int} v2);\\
\textbf{cons:}\\
    \>$\forall v:\textbf{int}.\texttt{LE}(v,v)$;\\
    \>$\forall v_1,v_2:\textbf{int}.(\texttt{LE}(v_1,v_2)\lor \texttt{LE}(v_2,v_1))$;\\
    \>$\forall v_1,v_2,v_3:\textbf{int}.(\texttt{LE}(v_1,v_2)\land \texttt{LE}(v_2,v_3)\Rightarrow \texttt{LE}(v_1,v_3))$;\\
\textbf{methods}:\\
    \>\textbf{int} compareTo(Comparable o);\\
    \>  \>\textbf{pre}\ \ \>\>$\rho\land o\neq \textbf{nil} \land \textbf{classOf}(o)=\textbf{theClass}$\\
    \>  \>\textbf{post}\  \>\>$\rho\land (\texttt{LE}(\texttt{VALUE}(),o{\fldacc}\texttt{VALUE}())\Rightarrow \textbf{ret} <=0) \land (\texttt{LE}(o{\fldacc}\texttt{VALUE}(),\texttt{VALUE}())\Rightarrow \textbf{ret}\ge 0)$\\
\}\\
\end{tabbing}
\end{boxedminipage}
\end{center}
\caption{The interface \texttt{Comparable}}\label{FIG-COMPARABLE}
\end{figure}

\subsection{Function symbols}
The function symbols declared in an interface are used to specify the methods of the interface.
For each symbol, the result type, arity, parameter types are given.
For each symbol $f$ declared in the interface, the memory scope function symbol, i.e. $\mathfrak{M}(f)$, is also implicitly declared in the interface.
$\mathfrak{M}(f)$ has the same formal parameters as $f$, and the return type of $\mathfrak{M}(f)$ is $\textbf{SetOf}(\textbf{Ptr})$.
Based on the rules for memory scopes given in \cite{DBLP:conf/ictac/ZhaoL13}, the definition of $\mathfrak{M}(f)$ in a class implementing the interface
can be constructed syntactically from the definition of $f$ in the same class.

There are two kinds of function symbols: \emph{class symbols} and \emph{object symbols}.
The object function symbols describe the properties about individual objects, while class symbols
describe the properties associated with the class.
The symbols declared with the keyword `\textbf{static}' are class symbols, and the rest are object symbols.
Given a function symbols $f$, $\mathfrak{M}(f)$ and $f$ are in the same kind.


Besides the explicitly declared function symbols, each interface has three additional special object function
symbols: \textbf{SetOf}(\textbf{Ptr}) \texttt{BLOCK}(), \textbf{SetOf}(\textbf{Ptr}) \texttt{pmem}(), and \textbf{bool} \texttt{INV}().
Intuitively speaking, \texttt{BLOCK} yields the memory units assigned to the member variables of the object,
\texttt{pmem}() yields the private memory owned by the object, and \texttt{INV}() is the invariant of the object.
It is required that the invariant of an object $o$, i.e. $o\fldacc \texttt{IVN}()$, holds before/after each method invocation to $o$.
For conciseness, the formula $\textbf{this}\fldacc \texttt{IVN}()$ is usually omitted in the pre-/post-conditions of the methods.

The function symbols declared with the keyword \textbf{attrib} are special object symbols called
\emph{attribute symbols}.
This kind of function symbols have no formal parameter. For each attribute symbol $f$ of an interface $I$, there is an implicit constraint template
$$\forall o:\textbf{theClass}. (o\neq \textbf{nil} \Rightarrow (o\fldacc\texttt{INV}()\Rightarrow o\fldacc\mathfrak{M}(f)()\subseteq \texttt{pmem}()))$$
Intuitively speaking, attribute symbols should only access the private memory of the object.
Both of the function symbols \texttt{pmem}() and \texttt{INV}() are attribute symbols.

In specifications, object function symbols are used in the form $e\fldacc f(\overline y)$, where $\overline y$ represents a list of suitable real parameters.
Usually, we abbreviate $\textbf{this}\fldacc f(\overline y)$ as $f(\overline y)$.
Class function symbols are used in the form $cexp\mbox{::}f(\overline y)$, where $cexp$ is either a class name, or the keyword \textbf{theClass}, or $\textbf{classOf}(e)$ for some expression $e$. Usually, we abbreviate $\textbf{theClass}\mbox{::}f(\overline y)$ as
$f(\overline y)$.

\begin{example}\label{FUN-SYM-DEC}
The interface \texttt{Comparable} in Fig.~\ref{FIG-COMPARABLE} declares two function symbols: the object symbol
\texttt{VALUE} and the class symbol \texttt{LE}. \texttt{VALUE} is an attribute symbol. The function symbol $\texttt{VALUE}$ yields an integer. The class function
(predicate) symbol $\texttt{LE}$ compares two integers.

The object function symbols $\texttt{INV}$ and $\texttt{pmem}$ are implicitly declared.
The memory scope function symbols $\mathfrak{M}(\texttt{VALUE})$ and $\mathfrak{M}(\texttt{LE})$ are also implicitly declared in \texttt{Comparable}.
Because \texttt{VALUE} is an attribute symbol, there is an implicit constraint template.
$$\forall o:\textbf{theClass}. (o\neq \textbf{nil} \Rightarrow (o\fldacc\texttt{INV}()\Rightarrow o\fldacc\mathfrak{M}(\texttt{VALUE})()\subseteq \texttt{pmem}()))$$
In the post-condition of the method \texttt{compareTo}, two occurrences of $\texttt{LE}$ are abbreviations for $\textbf{theClass}\mbox{::}\texttt{LE}$; the first
and the fourth occurrence of $\texttt{VALUE}()$ are abbreviations for $\textbf{this}\fldacc \texttt{VALUE}()$.
\hfill $\square$
\end{example}

\subsection{Constraint templates about the function symbols}
An interface also declares a set of constraint templates, which are a set of boolean expressions, about the declared function symbols.
Each class implementing the interface can give its own definitions to the function symbols declared in the interface. However, these definitions must satisfy
all of the constraints derived from these templates by substituting \texttt{theClass} with the class name. Based on these constraint templates, we can reason about formulas containing these function symbols
without knowing the exact definitions to which these symbols refer.

\begin{example}
The constraint templates declared in the interface \texttt{Comparable} show that $\texttt{LE}$ is a total order over integers.
Together with the attribute function $\texttt{VALUE}$, it indirectly casts a total order over the objects.

There is an implicit constraint template associated with each attribute symbol declared in \texttt{Comparable}. Such a constraint template about \texttt{VALUE} is already given
in Example~\ref{FUN-SYM-DEC}.
There are also similar constraint templates for \texttt{INV} and \texttt{pmem}.

For any class implementing \texttt{Comparable}, its definitions to \texttt{VALUE}, \texttt{INV}, \texttt{LE} and \texttt{pmem} must satisfy the constraints derived from these templates.
\hfill $\square$
\end{example}

\subsection{Method declarations and specifications}
For each method declared in an interface, the method name, the formal parameters, and the return type
are given. Besides these information, each method is specified using templates of pre-/post-conditions, which are formulas containing the keyword
\textbf{theClass}.
The function symbols declared in the interface can be used in these templates.
Suppose that $I$ is an interface, $m$ is a method declared in $I$,
we use $$I\mbox{::}m(\overline x) : \{P\}\_\{Q\}$$ to specify that the templates of pre-/post-conditions of $m$ are $P$ and $Q$ respectively.
For any class $C$ implementing $I$, the pre-/post-condition of $m$ defined in $C$ are derived by substituting \textbf{theClass} with $C$ in
$P$ and $Q$ respectively.

To reason about the relations between the values before/after method invocation, we can use $\overleftarrow e$ in the post-condition
to represent the value of $e$ on the pre-state.

\begin{example}
In Fig.~\ref{FIG-COMPARABLE}, the specification template of  \texttt{compareTo} includes the function symbols \texttt{VALUE} and \texttt{LE}.
Intuitively speaking, the specification says that the method returns negative, zero, or positive integers respectively
when \textbf{this} is less than, equal to, or greater than the parameter $o$.
Note that, all the occurrences  of \texttt{LE} in the specification template are the abbreviations for \textbf{theClass}::\texttt{LE}.

A class $C$ implementing \texttt{Comparable} can give different definitions to \texttt{VALUE} and \texttt{LE}, denoted as $C\mbox{::}\texttt{VALUE}$
and $C\mbox{::}\texttt{LE}$ respectively. Note that the two occurrences of $\texttt{LE}$ are abbreviations for $\textbf{theClass}\mbox{::}\texttt{LE}$,
the precondition and post-condition of \texttt{compareTo} in $C$ are respectively
$$\{\rho\land o\neq \textbf{nil} \land \textbf{classOf}(o)=C\}$$
and
$$\begin{array}{l}\{\rho\land (C\mbox{::}\texttt{LE}(\texttt{VALUE}(),o{\fldacc}\texttt{VALUE}())\Rightarrow \textbf{ret} \le 0) \land\\
\ \ \ \ \ \ \ (C\mbox{::}\texttt{LE}(o{\fldacc}\texttt{VALUE}(),\texttt{VALUE}())\Rightarrow \textbf{ret}\ge 0)\}\end{array}$$
\hfill$\square$
\end{example}

\section{The class definitions}\label{SEC-CLASS}
A class definition declares a set of member variables and defines a set of function symbols and methods. For simplicity,
we suppose that all the member variables are private (not accessible outside the class definition) and all the function symbols and methods are public (accessible outside the class definition).

A class definition also lists all the interfaces implemented by it.
For each interface implemented by this class, the following conditions hold.
\begin{itemize}
\item All the function symbols declared in the interface must be defined in this class.
    All the constraints derived from the constraint templates in the interface should be tautologies.
\item All the methods declared in the interface must be defined in the class. The specifications of these methods are derived by substituting \textbf{theClass} by the class name in the corresponding templates in the interface. 
\end{itemize}
A class definition can also define its own function symbols and methods. The preconditions and post-conditions of these methods must be given explicitly. There is one and only one method that has the same name with the class. This method is called the \emph{constructor} of the class, and is used to create objects of the class. A constructor has no return type.

The grammar of class definitions are depicted in Fig.~\ref{SYN-CLASS-DEFINITION}. The meta expression $(\dots)?$ means that the grammar component in the brackets is optional. In the grammar, `cName', `vName', `fName', `mName' are identifiers respectively representing class names, variable names, function names, and method names. The component `iList` lists the names of the interfaces implemented by this class. The other components are explained in the following subsections.

\begin{figure}
\begin{center}
\begin{minipage}{0.9\textwidth}
cDef ::= cName \textbf{impl} iList `\{' vDecs\ fDefs mDefs `\}'\\
vDecs ::= \textbf{var} `:' ( Type vName `;' )$^+$\\
fDefs ::= $\epsilon$ $|$ \textbf{func} `:' ( funcType Type fName (fPara) $\triangleq$ expressions `;' )$^+$\\
mDefs ::= \textbf{method} `:' ( mDef )$^+$\\
mDef  ::= ( Type )$?$ mName `(' fPara `)' ( \textbf{pre} Formula; \textbf{post} Formula; )$?$\\
\mbox{}\ \ \ \ \ \ \ \ \ \ \ \ \ \ \ \ \ `\{' vDecs statementList (\textbf{return} expressions;)$?$ `\}'
\end{minipage}
\end{center}
\caption{The syntax of class definitions}\label{SYN-CLASS-DEFINITION}
\end{figure}

\begin{example}
An example of the class definitions is given in Fig.~\ref{FIG-IMPLEMENTATION}. The class \texttt{Point} implements the interface \texttt{Comparable}.
Two member variables \texttt{x} and \texttt{y} are declared in this class. Both of the function symbols \texttt{VALUE} and \texttt{LE} declared in \texttt{Comparable} are defined in \texttt{Point}. This class also defines two attribute functions \texttt{FldX}() and \texttt{FldY}(), which yields the value of the member variables $\texttt{x}$ and $\texttt{y}$.
The method \texttt{compareTo} declared in \texttt{Comparable} is defined in \texttt{Point}. The specification of this method is
derived from the corresponding specification template in \texttt{Comparable} by substituting \textbf{theClass} with \texttt{Point}. For conciseness, this method specification is omitted in this class definitions.
The constructor of \texttt{Point} is defined in this class. It creates a new object with the member variables $\texttt{x}$ and $\texttt{y}$ set to 0. Three other methods, \texttt{Set}, \texttt{getX} and \texttt{getY}, are defined in this class. The specifications of these methods are explicitly given in this class.
\hfill$\Box$
\end{example}

\begin{figure}
\begin{center}
\begin{boxedminipage}{1.0\textwidth}
{\scriptsize
\begin{tabbing}
\ \ \ \ \=\ \ \ \ \ \=\ \ \ \ \ \ \ \ \=\ \ \ \ \=\\
\textbf{class} Point \textbf{impl} Comparable\\
\{\\
\textbf{var:}\\
    \>int \texttt{x};\\
    \>int \texttt{y};\\
\textbf{funcs:}\\
    \>\textbf{attrib} \textbf{int} $\texttt{VALUE}()\triangleq x*x+y*y$;\\
    \>\textbf{static} \textbf{bool} $\texttt{LE}(v_1,v_2)\triangleq v_1 \le v_2$;\\
    \>\textbf{attrib} \textbf{int} $\texttt{FldX}()\triangleq \texttt{x}$;\\
    \>\textbf{attrib} \textbf{int} $\texttt{FldY}()\triangleq \texttt{y}$;\\
\textbf{method}:\\
    \>Point() \textbf{pre} $\rho$ \textbf{post} $\rho\land(\mathfrak{M}(\rho)\cap \texttt{BLOCK}() = \emptyset)\land$ FldX()=0 $\land$ FldY()=0 \{x = 0; y = 0;\};\\
    \>\textbf{void} Set(int x1, int y1)\\
    \>  \>\textbf{pre} $\rho$ $\land$ ($\mathfrak{M}(\rho)$ $\cap$ pmem() = $\emptyset$) $\land$ true\\
    \>  \>\textbf{post} $\rho$ $\land$ (\texttt{FldX}()=x1 $\land$ \texttt{FldY}()=y1)\\
    \>  \>\{x:=x1; y:=y1;\};\\
    \>\textbf{int}  getX()  \textbf{pre} $\rho$ $\land$ \texttt{true} \textbf{post} $\rho$ $\land$ (\textbf{ret} = \texttt{FldX}()) \{\textbf{return} x;\};\\
    \>\textbf{int}  getY() \textbf{pre} $\rho$ $\land$ \texttt{true} \textbf{post} $\rho$ $\land$ (\textbf{ret} = \texttt{FldY}()) \{\textbf{return} y;\};\\
    \>\textbf{int} compareTo(Comparable* o)\\
    \>  \>\{\\
    \>  \>    \>\textbf{int} tmp1;\\
    \>  \>    \>\textbf{int} tmp2;\\
    \>  \>  \>tmp1 = o$\rightarrow$getX();\\
    \>  \>    \>tmp2 = o$\rightarrow$getY();\\
    \>  \>    \>return x*x + y*y - tmp1*tmp1 - tmp2*tmp2;\\
    \>  \>\}\\
\}\\
\end{tabbing}
}
\end{boxedminipage}
\end{center}
\caption{The class Point implementing the interface Comparable}\label{FIG-IMPLEMENTATION}
\end{figure}

\subsection{Member variable declarations}
The set of member variables are declared in the part `vDecs'. These member variables can only be accessed in the function symbol definitions and the method definitions of this class. In these definitions, a member variable $v$ can be accessed as \textbf{this}{\fldacc}$v$, or just $v$ for conciseness.

Member variables are private. They are a part of the implementation details of a class, so they should not be exposed to other class definitions. Because the method specifications will be used to reason about method-invocation statements outside the class definition, member variables can not occur in the method specifications.

\begin{example}
In the class definition in Fig.~\ref{FIG-IMPLEMENTATION}, two member variables \texttt{x} and \texttt{y} are declared. They can be directly accessed in the definitions of the function
symbols and methods in the \texttt{Point}. The keyword \textbf{this} is omitted in these definitions. For example, the definition of \texttt{VALUE} in \texttt{Point} is equivalent to
\begin{center}\textbf{attrib} \textbf{int} $\texttt{VALUE}()\triangleq \textbf{this}\fldacc x*\textbf{this}\fldacc x+\textbf{this}\fldacc y*\textbf{this}\fldacc y$;\\\end{center}
These member variables can not appear in the pre-/post-conditions of the methods. Instead, we use the function symbols \texttt{FldX}() and \texttt{FldY}() in the specifications.
\hfill$\Box$
\end{example}

\subsection{Function symbol definitions}
A set of function symbols are defined in the part `fDefs'. For each interface implemented by this class, all the function symbols
declared in the interface should be defined here.
The definition of the mandatory function BLOCK is derived directly from the part `vDecs'.
$$\textbf{attrib}\ \textbf{SetOf}(\textbf{ptr})\ \texttt{BLOCK}() \triangleq \{\textbf{this}\fldacc\&v | v\mbox{ is a member variable in `vDecs'}\}$$
People can define the mandatory functions \texttt{pmem}() and \texttt{INV}() in the class,
or just use the following default definitions.
\begin{center}
\begin{minipage}{0.6\textwidth}
$\textbf{attrib}\ \textbf{SetOf}(\textbf{ptr})\ \texttt{pmem}() \triangleq \textbf{this}\fldacc \texttt{BLOCK}()$\\
$\textbf{attrib}\ \textbf{bool}\ \texttt{INV}() \triangleq \textbf{true}$
\end{minipage}
\end{center}

The keyword \textbf{this} can not appear in the definitions of any class function symbols. Member variable access  is also forbidden in the definitions of class functions.

Given a function symbol $f$ defined in the class as
    $$f(\overline x)\triangleq e$$
the definition of $\mathfrak{M}(f)$ can be construct syntactically from the definition of $f$ following the rules given in \cite{DBLP:conf/ictac/ZhaoL13}. That is,
    $$\textbf{Setof}(\textbf{Ptr})\ \mathfrak{M}(f)(\overline x)\triangleq \mathfrak{M}(e)$$

The function symbol definitions in this class must satisfy all the constraints derived form the constraint templates  declared (explicitly or implicitly) in any interface implemented by the class.

\begin{example}
In Fig.~\ref{FIG-IMPLEMENTATION}, the class \texttt{Point} defines four function symbols: \texttt{VALUE}, \texttt{LE}, \texttt{FldX}, and \texttt{FldY}.
Among them, \texttt{VALUE} and \texttt{LE} are declared in the interface \texttt{Comparable}. The function \texttt{LE} is a class function symbol, so its definition does not access
member variables.

The functions \texttt{pmem} and \texttt{INV} are not explicitly defined in \texttt{Point}, so their definitions are
\begin{center}
\begin{minipage}{0.7\textwidth}
\textbf{attrib} \textbf{SetOf}(\textbf{ptr}) \texttt{pmem}() $\triangleq$ \{\&\textbf{this}\fldacc\texttt{x}, \&\textbf{this}\fldacc\texttt{y}\}\\
\textbf{attrib} \textbf{bool}    \texttt{INV}() $\triangleq$ \textbf{true}
\end{minipage}
\end{center}
%
The are three explicit constraint templates about \texttt{LE} in \texttt{Comparable}. Substituting \textbf{theClass} by \texttt{Point}, we have the following three constraints.
\begin{center}
\begin{minipage}{0.95\textwidth}
$\forall v:\textbf{int}.\texttt{Point}\mbox{::}\texttt{LE}(v,v)$;\\
$\forall v_1,v_2:\textbf{int}.(\texttt{Point}\mbox{::}\texttt{LE}(v_1,v_2)\lor \texttt{Point}\mbox{::}\texttt{LE}(v_2,v_1))$;\\
$\forall v_1,v_2,v_3:\textbf{int}.(\texttt{Point}\mbox{::}\texttt{LE}(v_1,v_2)\land \texttt{Point}\mbox{::}\texttt{LE}(v_2,v_3)\Rightarrow \texttt{Point}\mbox{::}\texttt{LE}(v_1,v_3))$;
\end{minipage}
\end{center}
It can be checked that the definition of $\texttt{Point}\mbox{::}\texttt{LE}$ satisfies all these constraints.

The following definition of $\mathfrak{M}(\texttt{VALUE})$ is derived from the definition of \texttt{VALUE} by the memory scope rule given in \cite{DBLP:conf/ictac/ZhaoL13}.
\begin{center}\textbf{attrib} \textbf{SetOf}(\textbf{Ptr}) $\mathfrak{M}(\texttt{VALUE})()\triangleq$\{\&\textbf{this}\fldacc x,\&\textbf{this}\fldacc y\};\end{center}
Because \texttt{VALUE} is an attribute symbol, the definition of $\mathfrak{M}(\texttt{VALUE})$ must satisfy the following constraint.
$$\forall o:\texttt{Point}. (o\neq \textbf{nil}\Rightarrow(o\fldacc \texttt{INV}()\Rightarrow(o\fldacc \mathfrak{M}(\texttt{VALUE})()\subseteq o\fldacc \texttt{pmem}()))) $$
From the definition of $\texttt{Point}\mbox{::}\texttt{pmem}$, this constraint is always satisfied.
\hfill$\Box$
\end{example}

\subsection{Method definitions}\label{Subsec-METHOD-DEF}
A set of methods are defined in the part `mDefs'. A class must define one and only one constructor, which is used to create new objects of this class.
The general grammar of method definitions is given in Fig.~\ref{SYN-CLASS-DEFINITION}. Given a method $m$ defined in a class $C$, we use the following formula
    $$C\mbox{::}m: \{P\}\_\{Q\}$$
to express that $P$ and $Q$ are respectively the pre-condition and post-condition of the method $m$ defined in $C$.

\subsubsection{Constructor definitions.}
Constructors are used to create new objects. The constructor of a class has the same name as the class.
A constructors do not return a value, so the return type and the `\textbf{return} expression' in the general grammar are absent. As a consequence, the keyword \textbf{ret} does not appear in the post conditions.
The object has not been created yet when a constructor is invoked, so the keyword \textbf{this} can not appear in the precondition of a constructor.

\subsubsection{Common method definitions.}
For each interface implemented by a class, all the methods declared in the interface should be defined in the class.
A class can also define its own methods.
If the return type of this method is \textbf{void}, the `\textbf{return} expression' in the general grammar is absent and
the keyword \textbf{ret} does not appear in the post-condition.

Let $m$ be a method declared in an interface $I$, and $C$ be a class implementing $I$.
The specification of $C\mbox{::}m$, i.e. the method defined in $C$, is derived by substituting $\textbf{theClass}$ with $C$ from the corresponding template declared in the interface.
That is, $I\mbox{::}m:\{P\}\_\,\{Q\}$ implies $C\mbox{::}m:\{P[C/\textbf{theClass}]\}\_\,\{Q[C/\textbf{theClass}]\}$.
For conciseness, the pre-/post-condition of $m$ are absent in the class definition in $C$.

\begin{example}
\label{EXAMPLE-METHOD-DEFINITION}
The class \texttt{Point} in Fig.~\ref{FIG-IMPLEMENTATION} defines the constructor of \texttt{Point}, which creates a new object with the fields \texttt{x} and \texttt{y} set to 0.
The method \texttt{compareTo} is declared in the interface \texttt{Comparable}. The specification of \texttt{compareTo} is not explicitly given in the class definition.
 It is derived by substituting \textbf{theClass} with \texttt{Point}
 in the template given in \texttt{Comparable}. So the precondition and post-condition are respectively
$$\{\rho\land o\neq \textbf{nil} \land \textbf{classOf}(o)=\texttt{Point}\}$$
and
$$\begin{array}{l}\{\rho\land (\texttt{Point}\mbox{::}\texttt{LE}(\texttt{VALUE}(),o{\fldacc}\texttt{VALUE}())\Rightarrow \textbf{ret} <=0) \land\\
\ \ \ \ \ \ \ (\texttt{Point}\mbox{::}\texttt{LE}(o{\fldacc}\texttt{VALUE}(),\texttt{VALUE}())\Rightarrow \textbf{ret}\ge 0)\}\end{array}$$
Three other methods are defined in \texttt{Point}: \texttt{getX}, \texttt{getY}, and \texttt{set}. They are used to access the member variables of the objects.
\hfill$\Box$
\end{example}

\subsection{Verifying a method w.r.t. its specification}
In this subsection, we discuss how to verify a method w.r.t. its specification.
For each method $m$ defined in a class $C$, if $m$ is declared in an interface implemented by $C$, its specification can be derived by a substitution in
the corresponding template; Otherwise, its specification is given in $C$ explicitly.
%

\subsubsection{Verification of constructor specifications.}
Let the constructor of a class $C$ be defined as
    $$C(\overline x)\ \textbf{pre}\ P;\ \textbf{post}\ Q;\ \{ \texttt{vDecs } \texttt{statement}\}$$
An execution of this constructor first allocates a new memory block for the new objects, and then executes the \texttt{statement}.
The formula $P$ holds when \texttt{statement} begins it execution if no local variable in \texttt{vDecs} occurs in $P$.
The keyword \textbf{this} refers to the newly allocated memory block for the class $C$ when \texttt{statement} begins its execution.
Let \textsf{LocMem} be the abbreviation for
    $$\{\&\texttt{v}\ |\mbox{\texttt{v} is a local variable decled in vDecs}\}$$
Let $\textsf{ExtraPre\_Constructor}$ be the abbreviation for
$$\textbf{this}\neq \textbf{nil} \land (\textbf{this}\fldacc \texttt{BLOCK}()\cap \mathfrak{M}(P)=\emptyset)\land (\textsf{LocMem}\cap \mathfrak{M}(P)=\emptyset)$$
We have that \textsf{ExtraPre\_Constructor} holds when \texttt{statement} begins it execution.

There may be some terms of the form $\overleftarrow e$ in the post-condition $Q$. The expression $\overleftarrow e$ means
the value of $e$ at the pre-state of \texttt{statement}. During the verification of the method body, $\overleftarrow e$ is equivalent to
$e@1$, where $1$ is the entry program point of the method body.
Let $Q'$ be a formula derived by substituting all the sub-expressions of the form $\overleftarrow e$ with $e@1$ in $Q$.
After the execution, the memory units for local variables are de-allocated. The formula $Q'$ is not affected by the de-allocation if no
local variable occurs in it.

The precondition $P$ and post-condition $Q$ may contain some assertion variables, which can be substituted with other formulas when
the specification of $C\mbox{::}m$ is used at the invocation place of $C\mbox{::}m$. However, the memory scopes of these formulas does not contains the memory units
allocated for the local variables because these memory units are unreachable at the invocation place. So the memory scopes
of $P$ and $Q$ is disjoint with \textsf{LocMem} if $P$ and $Q$ contains no local variable declared in \texttt{vDecs}.

Based on the above discussion, we have the following proof rule.
$$
\begin{array}{l}
\framebox{CSTOR}\frac
{
 \mbox{}\  \begin{array}{l}
        \{P\land \textsf{ExtraPre\_Constructor}\}
        \ \ \texttt{statement}\ \
        \{Q'\}
    \end{array}\
}
{
\begin{array}{c}
    C\mbox{::}C(\overline x) : \{P\}\_\{Q\}\\
    \mbox{\footnotesize $\ast$ $P$ and $Q$ contains no local variable declared in \texttt{vDecs}}\\
    \mbox{\footnotesize $\ast$ $Q'$ is derived by substituting all sub-expression $\overleftarrow e$ with $e@1$ in $Q$}
\end{array}
}\\
\end{array}
$$

\begin{example}
The symbols
\texttt{FldX} and \texttt{FldY} (abbreviations for \textbf{this}\fldacc\texttt{FldX} and \textbf{this}\fldacc\texttt{FldY} respectively)
in the specification of $\texttt{Point}\mbox{::}\texttt{Point}$ refer to their definitions in \texttt{Point}. It can be proved that the following specification holds.
$$\begin{array}{l}
\{\rho \land\textbf{this}\neq \textbf{nil} \land (\texttt{BLOCK}()\cap\mathfrak{M}(\rho)=\emptyset)\}\\
\mbox{}\ \ \ \ \ \ \ \ \ \ \ \texttt{x = 0; y = 0;}\\
\{\rho\land (\texttt{BLOCK}()\cap\mathfrak{M}(\rho)=\emptyset)\land \texttt{FldX}()=0 \land \texttt{FldY}()=0\}
\end{array}$$
No local variable is declared in this constructor. From the proof rule CSTOR, we prove the following specification.
$$\texttt{Point}\mbox{::}\texttt{Point}():\{\rho\}\_\,\{\rho\land (\texttt{BLOCK}()\cap\mathfrak{M}(\rho)=\emptyset)\land \texttt{FldX}()=0 \land \texttt{FldY}()=0\}$$
That is, the definition of \texttt{Point}\mbox{::}\texttt{Point} satisfies its specification.
\hfill$\Box$
\end{example}

\subsubsection{Verification of common method specifications.}
Suppose that the body of a method $m$ defined in the class $C$ is $\{ \texttt{vDecs; } \texttt{sts}; \}$,
and the precondition and post-condition are $P$ and $Q$ respectively.
When \texttt{sts} begins its execution, $P$ holds if it contains no local variable in \texttt{vDecs}; $\textbf{this}$ is
a non-nil pointer. The memory units for local variables are newly allocated.
Let \textsf{LocMem} be the abbreviation for
    $$\{\&\texttt{v}\ |\mbox{\texttt{v} is a local variable decled in vDecs}\}$$
We know that $(\mathfrak{M}(P)\cap \textsf{LocMem} = \emptyset)$ also holds when \texttt{sts} begins it execution.

Let $Q'$ be a formula derived by substituting all the sub-expressions of the form $\overleftarrow e$ with $e@1$ in $Q$.
After the execution of statement, the memory units allocated for local variables are de-allocated and the method returns.
If $Q'$ holds after the execution of \texttt{sts}, and no local variable occurs in $Q'$,
$Q$ holds after the invocation returns.

Based on the above discussion, we have the following proof rule.
$$
\begin{array}{l}
\framebox{METHOD}\frac
{
 \mbox{}\ \ \ \ \    \begin{array}{c}
        \{P\land \textbf{this}\neq \textbf{nil}\land (\mathfrak{M}(P)\cap \textsf{LocMem} = \emptyset)\}
         \ \ \texttt{sts}\ \
         \{Q'\}
    \end{array}\ \ \ \ \
}
{
\begin{array}{c}C\mbox{::}m(\overline x) : \{P\}\_\{Q\}\\
\mbox{\footnotesize $\ast$ $P$ and $Q$ contains no local variable declared in \texttt{vDecs}}\\
\mbox{\footnotesize $\ast$ $Q'$ is derived by substituting all sub-expression $\overleftarrow e$ with $e@1$ in Q}
\end{array}
}
\end{array}
$$

\begin{example}
The method \texttt{compareTo} is declared in \texttt{Comparable}. Its pre-/post-conditions are given in Example~\ref{EXAMPLE-METHOD-DEFINITION}.
Let \textsf{compareTo\_Pre} and \textsf{compareTo\_Post} are respectively abbreviations for these two formulas. The specification
of \texttt{compareTo} can be written as
$$\texttt{Point}\mbox{::}\texttt{compareTo}(o):\{\textsf{compareTo\_Pre}\}\_\,\{\textsf{compareTo\_Post}\}$$
There is no sub-expressions of the form $\overleftarrow e$ in \textsf{comparTo\_Post}.
According to the proof rule METHOD, this specification holds if the following specification holds.
$$
\begin{array}{l}
\{\textsf{compareTo\_Pre}\land (\{\&\texttt{tmp1},\&\texttt{tmp2}\}\cap (\mathfrak{M}(\rho))=\emptyset\}\\
\mbox{\ \ \ \ \ \ \ \ \ \ The body of \texttt{Point}::\texttt{compareTo}}\\
\{\textsf{compareTo\_Post}\}
\end{array}
$$
\hfill$\Box$
\end{example}

\section{Types and expressions}\label{SEC-TYPE-EXPRESSION}
In this subsection, we discuss the types, expressions, and statements associated with interfaces and classes.

\subsection{Types}
The small language used in this paper is slightly different from the one used in \cite{DBLP:conf/ictac/ZhaoL13}.
It supports \textbf{int}, \textbf{bool}, array types, struct types, pointer types, interface types, and class types.
\begin{itemize}
\item An interface $I$ declared in the program is a type. The value of an expression with a static type $I$ is a reference to
    an object of some class implementing $I$.
\item A class $C$ defined in the program is also a type. The value of an expression with type $C$ is a reference to
    an object of the class $C$.
\end{itemize}
One memory unit is assigned to each variable declared with a class type or an
interface type in the runtime. The member variables of an object is stored in a memory block.
The memory layout for objects are same as the memory layout for record types in \cite{DBLP:conf/ictac/ZhaoL13}.
 In memory assignment, a variable with type $I$ or $C$ is treated as a pointer to a record type, and the member variables are treated as fields of the record type.
So we can still use the axioms for memory layout of record types in that paper.

If $C$ is a class implementing $I$, $I$ is a super type of $C$. That is, a $C$ object can be assigned to a variable with a static type $I$.
Given an expression $e$ with static type $I$, the value of $e$ in the runtime is either \textbf{nil} or refers to an object of some class $C'$
implementing $I$. We call $C'$ the \emph{dynamic class} of $e$, denoted as $\textbf{classOf}(e)$.

Because inheritance is not supported in the small programming language in this paper, given an expression $e$ with static type $C$, the value of $e$ in the runtime is either \textbf{nil}, or refers to an object of $C$.

\subsection{Expressions associated with interfaces and classes.}
There are five kinds of expressions associated with interfaces used in programs.
\begin{itemize}
\item \textbf{this.} It is a keyword used in interface declarations and class definitions. It refers to the current object being manipulated.
    This keyword can not appear in the constraints, constraints templates,
    and the definitions of class function symbols.
    The static type of \textbf{this} is $I$ (or $C$ ) when it appears in the declaration of an interface $I$ (or the definition of a class $C$).
\item \textbf{Member variables.} In our language, the member variables declared in classes are always private.
     A member variables $v$ in a class $C$ can only occur in the method bodies and object-function symbol definitions
    in $C$, in the form $\textbf{this}\fldacc v$. Usually, this expression is abbreviated as $v$ for conciseness. The static type of this expression is just the declared type of $v$. This expression has a left-value, i.e. a value can be assigned to $\textbf{this}\fldacc v$.
\item \textbf{ret.} It is a keyword used only in the post-conditions of methods. It represents the return value of the method. The static type of \textbf{ret} is the return type of the method.
\item \textbf{Object function symbol calls.} Let $f$ be an object function symbol declared as
                $$T\ f(\overline x)$$
    in an interface $I$ (or defined in a class $C$), $e\fldacc f(\overline y)$ is an expression with static type $T$ if the following conditions hold.
    \begin{itemize}
    \item The static type of $e$ is $I$ (or $C$ respectively).
    \item The real parameters $\overline y$ conform to the formal parameters $\overline x$ of $f$.
    \end{itemize}
    This expression calls the definition of $f$ in the class $\textbf{classOf}(e)$.
    For conciseness, the expression $\textbf{this}\fldacc f(\overline y)$ is usually abbreviated as $f(\overline y)$.
\item \textbf{Class function symbol calls.} Let $f$ be a class function symbol declared as
                $$T\ f(\overline x)$$
    in an interface $I$ (or defined in a class $C$), $cexp\mbox{::}f(\overline y)$ is an expression with static type $T$ if the following conditions hold.
    \begin{itemize}
    \item The real parameters $\overline y$ conform to the formal parameter $\overline x$ of $f$;
    \item The class-prefix $cexp$ is either a class name, or the keyword \textbf{theClass}, or $\textbf{classOf}(e)$ for some expression $e$ with static type $I$ (or $C$ respectively).  If $cexp$ is a class name, the class must implement $I$ (or just $C$).
    \end{itemize}
    This expression calls the definition of $f$ in the class denoted by $cexp$.
    For conciseness, the expression $\textbf{theClass}\mbox{::}f(\overline y)$ in an interface declaration is usually abbreviated as $f(\overline y)$. In a class definition $C$, $C\mbox{::}f(\overline y)$ is usually abbreviated as $f(\overline y)$.
\end{itemize}
The memory scope rules for these expressions are given in Table~\ref{TAB-MEMORY-SCOPE}.
\begin{table}
\begin{center}
\begin{tabular}{|c|c|}
  \hline
  The expressions & The memory scopes\\
  \hline
  a class name & $\emptyset$ \\
  \hline
  $\textbf{theClass}$ & $\emptyset$ \\
  \hline
  $\textbf{classOf}(e)$ & $\mathfrak{M}(e)$ \\
  \hline
  $\textbf{this}$ & $\emptyset$ \\
  \hline
  $\textbf{this}\fldacc v$ & $\&\textbf{this}\fldacc v$\\
  \hline
  $e\fldacc f(\overline y)$ & $\mathfrak{M}(e)\cup \mathfrak{M}(\overline y)\cup e\fldacc\mathfrak{M}(f)(\overline y)$ \\
  \hline
  $cexp\mbox{::}f(\overline y)$  & $\mathfrak{M}(cexp)\cup \mathfrak{M}(\overline y)\cup cexp\mbox{::}\mathfrak{M}(f)(\overline y)$ \\
  \hline
\end{tabular}
\end{center}
\caption{The memory scopes of expressions associated with interfaces and classes}\label{TAB-MEMORY-SCOPE}
\end{table}

\section{Statements and their proof rules}\label{SEC-STATEMENT}
In the small language used in this paper, the \textbf{alloc} statement is no longer supported.
Instead, people can use the object creation statement to create objects. Three new kinds of statements is discussed in this subsection : the return statements, the object creation statements, and the method invocation statements.

\subsection{The return statements.}
The \textbf{return} statement is the last statement of a method definition. The statement `\textbf{return} $exp$' first evaluates the value of $exp$, and then returns
this value. So we have the following axiom.
$$
\textrm{RETURN-ST: }
     \begin{array}{l}
            \mbox{\ \ \ \ \ \ \ }\{Q[exp/\textbf{ret}]\}\ \ \ \textbf{return\ } exp\ \ \ \{Q\}
            \end{array}
$$

\subsection{Object creation statements.}
Given a class $C$, the following statement
        $$e_0:=\textbf{new}\ C(\overline y)$$
first evaluates the left-value of $e_0$, (i.e. $\&e_0$), and then creates a new object using the real parameter $\overline y$,
and finally stores the reference into the memory unit referred by $\&e_0$. The left-hand $e_0$ must have a left-value and the
static type of $e_0$ is either $C$, or an interface $I$ implemented by $C$. The real-parameters $\overline y$ conforms to the
formal parameters of the constructor of $C$.

When this statement begins its execution, the precondition of the constructor must be satisified.
After the execution,
the formula derived by substituting \textbf{this} and formal parameters with the object reference and real parameters respectively in post-condition of the constructor of $C$ holds.  So we have the following proof rule.
    $$
\framebox{OBJ-CREATION}
\frac{ \textrm{cName}\mbox{::}\textrm{cName}(\overline x) : \{P\}\_\{Q\}}
        {\mbox{\ \ \ \ \ \ \ }
            \begin{array}{l}
            \{i:(\&e_0\neq \textbf{nil})\land P[\overline y/\overline x]\}\\
             \mbox{\ \ \ \ \ \ \ }e_0 =\textbf{new}\ \textrm{cName}(\overline y)\\
            \{j:(\ast(\&e_0@i)\neq \textbf{nil})\land Q[\ast(\&e_0@i)/\textbf{this}][\overline y@i/\overline x]\}
            \end{array}
            \mbox{\ \ \ \ \ \ \ }
        }
$$

\subsection{Class method invocation statements.}
A method $m$ defined in a class $C$ can be invoked as  $e_0:=e_1{\fldacc}\textrm{m}(\overline y)$ if the return type of $m$ is not \textbf{void};
or otherwise invoked as $e_1{\fldacc}\textrm{m}(\overline y)$.
It is required that $e_1$ is an expression with static type $C$ and the real parameters $\overline y$ must conform to the formal parameters of $m$.
For the first form, $e_0$ must have a left-value and the type of $e_0$ is a super type of the return type of $m$.

The statement $e_0:=e_1{\fldacc}\textrm{m}(\overline y)$ invokes the method $C\mbox{::}m$, i.e. the method $m$ defined in the class $C$.
Let $C\mbox{::}m:\{P\}\_\,\{Q\}$ be the specification about $C\mbox{::}m$.
Before the invocation,  $e_1$ must refer to an object of $C$, and $\&e_0$ must be a non-nil pointer. Furthermore,
the precondition $P[\overline y/\overline x][e_1/\textbf{this}]$ must hold. After the invocation, the property
$Q[\ast(\&e_0@i)/\textbf{ret}][\overline y@i/\overline x][e_1@i/\textbf{this}]$ holds.
So we have the following proof rule for the class method invocations of the first form.
$$
\framebox{C-INVOC-1}\frac{C\mbox{::}m(\overline x) : \{P\}\_\{Q\}}
{
 \mbox{}\ \ \ \ \    \begin{array}{l}
        \{i:(\&e_0\neq \textbf{nil})\land (e_1\neq \textbf{nil})\land  P[\overline y/\overline x][e_1/\textbf{this}]\}\\
        \ \ \ \ \ \ \ \ \ \ e_0=e_1\fldacc m(\overline y)\\
        \{j:Q[\ast(\&e_0@i)/\textbf{ret}][\overline y@i/\overline x][e_1@i/\textbf{this}]\}
    \end{array}\ \ \ \ \
}
$$
Similarly, we have the proof rule for the class method invocations of the second form.
$$
\framebox{C-INVOC-2}\frac{C\mbox{::}m(\overline x) : \{P\}\_\{Q\}}
{
 \mbox{}\ \ \ \ \    \begin{array}{l}
        \{i: (e_1\neq \textbf{nil})\land P[\overline y/\overline x][e_1/\textbf{this}]\}\\
        \ \ \ \ \ \ \ \ \ \ e_1\fldacc m(\overline y)\\
        \{j:(Q[\overline y@i/\overline x][e_1@i/\textbf{this}]\}
    \end{array}\ \ \ \ \
}
$$

\subsection{Interface method invocation statements.}
A method $m$ declared in an interface $I$ can be invoked as  $e_0:=e_1{\fldacc}\textrm{m}(\overline y)$ if the return type of $m$ is not \textbf{void};
or invoked as $e_1{\fldacc}\textrm{m}(\overline y)$ otherwise.
It is required that $e_1$ is an expression with static type $I$ and the real parameters $\overline y$ must conform to the formal parameters of $m$.
For the first form, $e_0$ must have a left-value and the type of $e_0$ is a super type of the return type of $m$.

For conciseness, we usually use $e_0:= m(\overline y)$ or $m(\overline y)$ as abbreviations for $e_0:=\textbf{this}{\fldacc}m(\overline y)$ or
$\textbf{this}{\fldacc}m(\overline y)$ respectively.

The statement  $e_0:=e_1{\fldacc}\textrm{m}(\overline y)$ invokes the method $m$ defined in the class $\textbf{classOf}(e_1)$ by the real-parameters $\overline y$, and store
the return value in the memory unit $\&e_0$. Suppose that $I\mbox{::}m:\{P\}\_\{Q\}$ holds, and $C_1,C_2\dots,C_n$ are all the classes implementing $I$.
The dynamic class of $e_1$, i.e. $\textbf{classOf}(e_1)$, is  $C_k$ for some $k$ $(1\le k\le n)$ if $e_1\neq \textbf{nil}$.
This statement invokes $C_k\mbox{::}m$ when the dynamic class of $e_1$ is $C_k$.
According to the discussion about method specification in Subsection~\ref{Subsec-METHOD-DEF},
the precondition of the method $C_k\mbox{::}m$ is $P[C_k/\textbf{theClass}]$. From the discussion about class invocations,
the precondition of the statement is $e_1\neq \textbf{nil}$ and
$$\begin{array}{rl}
        & (\textbf{classOf}(e_1)= C_1) ?  P[C_1/\textbf{theClass}][\overline y/\overline x][e_1/\textbf{this}]\\
    :\  & (\textbf{classOf}(e_1)= C_2) ?  P[C_2/\textbf{theClass}][\overline y/\overline x][e_1/\textbf{this}]\\
    :\  &\ \ \ $\dots\ \ \ \ \dots$\\
        & (\textbf{classOf}(e_1)= C_n) ?  P[C_n/\textbf{theClass}][\overline y/\overline x][e_1/\textbf{this}]\\
    :\  &\textbf{false}
\end{array}
$$
The above conditional formula is equivalent to $$P[\textbf{classOf}(e_1)/\textbf{theClass}][\overline y/\overline x][e_1/\textbf{this}]$$
Let $i$ be the program point before this invocation statement, the post-condition of the statement $e_0:=e_1{\fldacc}\textrm{m}(\overline y)$ is
$$Q[\textbf{classOf}(e_1@i)/\textbf{theClass}][\ast(\&e_0@i)/\textbf{ret}][\overline y@i/\overline x][e_1@i/\textbf{this}]$$
So we have the following proof rule.
$$
\framebox{I-INVOC-1}\frac{I\mbox{::}m(\overline x) : \{P\}\_\{Q\}}
{
 \mbox{}\ \ \ \ \    \begin{array}{l}
        \{i:(\&e_0\neq \textbf{nil})\land (e_1\neq \textbf{nil})\land\\
        \ \ \ \ \  P[\textbf{classOf}(e_1)/\textbf{theClass}][\overline y/\overline x][e_1/\textbf{this}]\}\\
        \ \ \ \ \ \ \ \ \ \ e_0=e_1\fldacc m(\overline y)\\
        \{j:Q[\textbf{classOf}(e_1@i)/\textbf{theClass}][\overline y@i/\overline x]\\
        \ \ \ \ \ \ \ \ [e_1@i/\textbf{this}][\ast(\&e_0@i)/\textbf{ret}]\}
    \end{array}\ \ \ \ \
}
$$
For the interface method invocation statements of the form $e_1\fldacc m(\overline y)$, we can similarly get the following proof rule.
$$
\framebox{I-INVOC-2}\frac{I\mbox{::}m(\overline x) : \{P\}\_\{Q\}}
{
 \mbox{}\ \ \begin{array}{l}
        \{i: (e_1\neq \textbf{nil})\land P[\textbf{classOf}(e_1)/\textbf{theClass}][\overline y/\overline x][e_1/\textbf{this}]\}\\
        \ \ \ \ \ \ \ \ \ \ e_1\fldacc m(\overline y)\\
        \{j:Q[\textbf{classOf}(e_1@i)/\textbf{theClass}][\overline y@i/\overline x][e_1@i/\textbf{this}]\}
    \end{array}\
}
$$

\section{Code verification under the open-world assumption}\label{SEC-OPEN-WORLD}
The proof rules I-INVOC-1 and I-INVOC-2 show that we can have the preconditions and post-conditions of the interface method invocation statements without
knowing the exact dynamic class of the receivers. These preconditions and post-conditions usually contain terms of the form $\textbf{classOf}(e)$, or
$\textbf{classOf}(e)\mbox{::}f(\overline y)$.
Next, we will discuss how to deal with such terms.

If the dynamic class of an expression $e$ is $C$, we know that for any formula template $P$, $P[C/\textbf{theClass}]$ is equivalent
to $P[\textbf{classOf}(e)/\textbf{theClass}]$.
%
Let $I$ be an interface, $C_1,C_2,\dots,C_n$ be all the classes implementing $I$, and $constr$ be a constraint template declared in $I$.
According to the proof obligations of class definitions, $constr[C_k/\textbf{theClass}]$ holds for each $k(1\le k\le n)$. Given an expression $e$ with static type $I$,
the dynamic class of $e$, i.e. $\textbf{classOf}(e)$, must be some class implementing $I$ if $e$ is not \textbf{nil}. So the following formula holds and can be used in
the code verification.
$$(e\neq \textbf{nil})\Rightarrow constr[\textbf{classOf}(e)/\textbf{theClass}] $$
This formula holds even if some new classes implementing $I$ are added into the program, because the function symbol definitions in these new classes must also satisfy
the constraints.

\begin{figure}
\begin{boxedminipage}{0.9\textwidth}
{\scriptsize
\begin{tabbing}
\mbox{}\ \ \ \ \ \=\ \ \ \ \ \=\ \ \ \ \ \=\ \ \ \ \ \=\ \ \ \ \ \=\ \ \ \ \ \=\\
class arrayList\{\\
\textbf{var}:\\
    \>\texttt{Comparable} \verb"arr"[10];\\
\textbf{funcs}:\\
    \>\textbf{SetOf}(\textbf{Ptr}) \texttt{pmem}() $\triangleq$ $\lambda$x.(\&arr[x])[0..9];\\
    \>\texttt{Comparable} \texttt{get}(int i) $\triangleq$ \verb"arr"[i];\\
    \>\textbf{bool} \texttt{MemLayOut}() $\triangleq$ $\forall$i$\in$(0..9)($\forall$j$\in$(0..9)(\&arr[j] $\not\in$ get(i){\fldacc}\texttt{pmem}()))\\
\textbf{method}:\\
    \>\textbf{void} Set(\texttt{Comparable} obj, int i)\\
    \>  \>\textbf{Pre} \>\>$\rho\land (\mathfrak{M}(\rho) \cap \texttt{pmem}() = \emptyset) \land 0 \le i \land i \le 9$\\
    \>  \>\textbf{Post} \>\>$\rho\land (\texttt{get}(i) = \texttt{obj}) \land \forall x\in (0..9)((x \neq i) \Rightarrow (\texttt{get}(x) = \overleftarrow{\mbox{\texttt{get}(x)}}))$\\
    \>  \>\{arr[i] = obj;\}\\
    \\
    \>\textbf{void} Sort()\\
    \>  \>\textbf{Pre} \>\>$\rho\land (\mathfrak{M}(\rho) \cap \texttt{pmem}() = \emptyset) \land {\texttt{MemoryLayout}}() \land$\\
    \>  \>              \>\>$(\forall x\in(0..9)(\texttt{get}(i) \neq \textbf{nil})) \land (\forall x\in (0..9)(\textbf{classOf}(\texttt{get}(i))=\textbf{classOf}(\texttt{get}(0))))$\\
    \>  \>\textbf{Post}\>\>$\rho\land (\textbf{classOf}(\texttt{get}(0)) = \textbf{classOf}(\overleftarrow{\texttt{get}(0)}))\land$\\
    \>  \>              \>\>$\forall i\in (0..8)(\textbf{classOf}(\texttt{get}(0))::\texttt{LE}( \texttt{get}(i){\fldacc}\texttt{VALUE}(), \texttt{get}(i+1){\fldacc}\texttt{VALUE}()))$\\
\>     \> \{\\
\>     \>  \>int i,j,cR;\\
\>     \>  \>Points *tmp;\\
\>     \>  \>i = 9;\\
\>     \>  \>while(i$>$0)\{\\
\>     \>  \>    \>j = 0;\\
\>     \>  \>    \>while (j$<$i-1)\\
\>     \>  \>    \>\{\>cR = arr[j]$\rightarrow$compareTo((classOf(arr[j]))arr[j+1]);\\
\>     \>  \>    \>  \>if(cR $>$ 0)\{\\
\>     \>  \>    \>  \>  \>temp = arr[j];  arr[j]=arr[j+1];  arr[j+1]=temp;\\
\>     \>  \>    \>  \>\}\\
\>     \>  \>    \>  \>else\\
\>     \>  \>    \>  \>  \>skip;\\
\>     \>  \>    \>  \>j = j+1;\\
\>     \>  \>    \>\}\\
\>     \>  \>\}\\
\>     \>  \>i = i-1;\\
\>     \>  \}\\
\}\\
\end{tabbing}
}
{\small //The code using ArraySort}\\
\scriptsize
ArrayList* al = new ArraySrot()\\
//Add ten Point objects into the object al.\\
$\dots \dots$;\\
$\dots \dots$;\\
al{\fldacc}Sort();\\
\end{boxedminipage}
\caption{The sort algorithm for Comparable objects}\label{COMPARABLE-SORT}
\end{figure}

\begin{example}
Part of the class \texttt{arrayList} is given in Fig.~\ref{COMPARABLE-SORT}. An \texttt{arrayList} object stores some \texttt{Comparable} objects in the member variable \verb"arr".
The specification of the method \texttt{sort} is given in the class definition. If all the elements of $\verb"arr"$ are not \textbf{nil} and point to objects of the same class,
the method \texttt{sort} can sort these objects w.r.t. to the order induced by the method \texttt{compareTo}.

Because of the space limitation, we just briefly show how to prove that the following formula is an invariant of the inner while-statement.
\begin{equation}
\begin{array}{l}
\forall x\in (0..j-1) (\textbf{classOf}(\verb"arr"[0]@1)\mbox{::}\texttt{LE}(\\
\mbox{}\ \ \ \ \ \ \ \ \ \ \ \ \ \ \ \ \ \ \ \ \ \ \ \ \ \ \ \ \ \verb"arr"[x]\fldacc\texttt{VALUE}(),\verb"arr"[j]\fldacc\texttt{VALUE}())) \land \\
\forall x \in(0..9)(\verb"arr"[x] \neq\textbf{nil}) \land\\
\forall x\in (0..9) (\textbf{classOf}(\verb"arr"[x])=\textbf{classOf}(\verb"arr"[0]@1))
\end{array}\label{INV-EXAMP}
\end{equation}
Now we prove that Formula~\ref{INV-EXAMP} holds at the end of the loop-body if it holds at the beginning of the loop-body.

Because $\&\texttt{cR}$ is not in the memory scope of Formula \ref{INV-EXAMP},
Formula~\ref{INV-EXAMP} still holds at the program point after the assignment to \texttt{cR}, i.e. the point before the if-statement.
Furthermore, based on the specification of \texttt{compareTo} and the proof rule \textrm{I-INVOC-1}, the following formula also holds
at the point after the assignment to \texttt{cR}.
\begin{equation}
\begin{array}{l}
(\textbf{classOf}(\verb"arr"[\mbox{j}])\mbox{::}\texttt{LE}(\verb"arr"[\mbox{j}]\fldacc\texttt{VAL}(),\verb"arr"[\mbox{j+1}]\fldacc\texttt{VAL}())\Rightarrow \mbox{cR} \le 0) \land\\
(\textbf{classOf}(\verb"arr"[\mbox{j}])\mbox{::}\texttt{LE}(\verb"arr"[\mbox{j+1}]\fldacc\texttt{VAL}(),\verb"arr"[\mbox{j}]\fldacc\texttt{VAL}())\Rightarrow \mbox{cR} \ge 0)
\end{array}\label{RELATION}
\end{equation}
Let $\textbf{CLS}$ be an abbreviation for $\textbf{classOf}(\verb"arr"[0]@1)$. Because $\verb"arr"[0]@1\neq \textbf{nil}$ holds, substituting \textbf{theClass} with $\textbf{classOf}(\verb"arr"[0]@1)$ in the constraint templates declared in \texttt{Comparable},  we have
  $$\begin{array}{l}
  \forall v_1,v_2,v_3:\textbf{int}.(\textbf{CLS}\mbox{::}\texttt{LE}(v_1,v_2)\land \textbf{CLS}\mbox{::}\texttt{LE}(v_2,v_3)\Rightarrow\textbf{CLS}\mbox{::}\texttt{LE}(v_1,v_3))\\
    \forall v_1,v_2:\textbf{int}.(\textbf{CLS}\mbox{::}\texttt{LE}(v_1,v_2)\lor \textbf{CLS}\mbox{::}\texttt{LE}(v_2,v_1))
    \end{array}$$
From these constraints, it can be implied from Formula~\ref{INV-EXAMP} and \ref{RELATION} that
$$\begin{array}{l}
    (\mbox{cR} \le 0)?\\
    \mbox{}\ \, \forall x\in (0..\mbox{j}+1) (\textbf{CLS}\mbox{::}\texttt{LE}(\verb"arr"[x]\fldacc\texttt{VAL}(),\verb"arr"[\mbox{j}+1]\fldacc\texttt{VAL}()))\\
    :\forall x\in (0..\mbox{j}+1) (\textbf{CLS}\mbox{::}\texttt{LE}(\verb"arr"[x]\fldacc\texttt{VAL}(),\verb"arr"[\mbox{j}]\fldacc\texttt{VAL}()))
    \end{array}
$$
Using the weakest-precondition computation algorithm presented in \cite{DBLP:conf/ictac/ZhaoL13}, it can be proved that the invariant holds at the end of the loop body.

A piece of client code using this class is given following the definition of \texttt{arrayList}. An \texttt{arrayList} object \texttt{al} is first created
and then ten \texttt{Point} objects are set into this object.
The following condition holds at the point before the statement al{\fldacc}Sort().
\begin{center}
\begin{minipage}{0.9\textwidth}
$\forall$x$\in$(0..9)(\texttt{al}{\fldacc}\texttt{get}(i) $\neq$ \textbf{nil})) $\land$ \texttt{al}{\fldacc}\texttt{MemoryLayout}() $\land$\\
$\forall$x$\in$(0..9)(\textbf{classOf}(al{\fldacc}\texttt{get}(i)) = \texttt{Point})
\end{minipage}
\end{center}
So after this statement, the following property holds.
\begin{center}
$\forall$i$\in$(0..8)(\texttt{Point}::\texttt{LE}(al{\fldacc}get(i){\fldacc}\texttt{VALUE}(), al{\fldacc}get(i+1){\fldacc}\texttt{VALUE}()))
\end{center}
\hfill$\Box$
\end{example}

\section{Some discussions on encapsulation and inheritance}\label{SEC-DISCUSSION}

\subsection{About the \texttt{pmem} and \texttt{INV} functions}\label{PMEM-INV}
The function symbols \texttt{pmem} and \texttt{INV} are important in the approach presented in this paper.

Given an object $o$, the expression $o\fldacc \texttt{pmem}()$ represents the private memory owned by this object.
For any attribute symbol \texttt{att} of $C$, the value of $o\fldacc \texttt{att}()$ keeps unchanged if no memory unit in $o\fldacc \texttt{pmem}()$ is changed.
A good programming practice is to make each object manage it own private memory and do not share them with other objects.
Such features can be checked through the escape analysis technique\cite{ESCAPE-ANALYSIS}.
For such an object $o$, we have the following conclusions.
\begin{itemize}
\item A method invocation $o\fldacc m$ may modify the private memory of $o$. The properties about the attributes of $o$ can be derived from the specification of $m$;
\item For other statements, the private memory of $o$ keeps unchanged, so all the attribute values of $o$ keep unchanged.
\end{itemize}
From these conclusions, we can reason about the attribute values without referring to the complicated internal structure of the objects.
The private memory of such an object can be managed as a whole during the verification.

The function symbol \texttt{INV} is an attribute one, and denotes the invariant of an object.
This invariant must hold before/after each method invocation. If an object does not share its private memory with other objects, it is easy to
check that the invariant of this object always holds before/after any method invocation.
Subclasses can have stronger invariants than their super-classes. So a subclass can define an overriding method that has a stronger invariant as a part of
its precondition, while the client code don't have to be re-verified.

Note that, an object may be designed to share its private memory with others for some reasons. For example, a Container object in Java shares
its private memory with its Iterator objects. In such cases, we may have to treat the invariants carefully. It is important to make sure that
a method of an object does not break the invariants of another related object.

\subsection{About class inheritance}
For simplicity, we only discuss  in this paper how to deal with the implementation relation between interfaces and classes.
However, the same idea can also be applied to class inheritance.

In an ordinary object-oriented language, the definition of a class $B$ can be viewed from two aspects.
\begin{itemize}
\item The interface of $B$, which specifies how the class can be used. Here, the interface means only the set of method signatures of $B$.
\item The implementations of $B$, i.e. the implementations of the method signatures of $B$.
\end{itemize}
A sub-class $D$ of $B$ inherits (and may extends) the interface of $B$. It overrides ( i.e. gives new implementations to )
some methods of $B$ and inherits the others. If people want $D$ be a subtype of $B$, i.e.
a $D$ object can be used in the places a $B$ object is used, some restrictions on the new implementations are required.

Now we discuss two slightly different cases. The first, a superclass is known to be extended by some subclasses at the design time;
The second, a new subclass is added after the superclass has been finished. The second case usually happens in the software maintenance phase.

In the following discussion, it is assumed that all the methods are specified by function symbols defined
in their classes. No member variable is referred directly in the specification.
\subsubsection{The case a superclass is designed to be inherited.} In this case, programmers should know how the super class is used. So they can give
a set of properties about the function symbols defined in the superclass $B$ such that the following conditions hold.
\begin{itemize}
\item The properties hold w.r.t. the function symbol definitions given in $B$.
\item These properties are sufficient to verify the client code using the superclass $B$. The symbol
definitions in $B$ are not directly used in the verification.
\end{itemize}
Thus, the signatures of function symbols and methods, together with these properties, compose a complete interface specifications of $B$.
Programmers can define the subclass $D$ by implementing this interface specification as described in the previous sections: some function symbols
and method are inherited, and the others are overridden. All the constraints (properties) given in $B$ must be satisfied.

It is not necessary to re-verify a method inherited from $B$ if non of the function symbols occurred in its specifications are overridden in $D$.
Otherwise, the method must be re-verified w.r.t. the new function symbol definitions in $D$.

Roughly speaking, we can view the class definition of $B$ as an interface specification plus a class definition in this case.

\subsubsection{To design a subclass when the superclass and the client code has been finished.}\label{SUBSUB-INHERITANCE-CLIENT-CODE-FIXED}
As the interface specification of the superclass is not given, the definitions of function symbols in the superclass are
used directly to verify the client code using this superclass.

To implement a subclass such that it can substitute the superclass in the client code, the interface specification of the superclass
must be reconstructed.
When function symbol definitions are used during the verification of the client code, people in fact use some properties
implied by these function symbol definitions. By investigating the verification, all of such properties
can be retrieved. These properties, together with the function symbols and the method specifications, compose an interface
specification. If a subclass implements this interface, it can substitute the superclass without re-verifying the client code.

\subsubsection{To design a subclass which can substitutes the superclass in any client code.}
Suppose that the predicate symbol \texttt{INV} are defined as $\texttt{INV}_{sup}$ and $\texttt{INV}_{sub}$ in the superclass and the subclass respectively.
For each method $m$, the pre-condition and post-condition of $m$
are $\texttt{INV}_{sup}\land P_{m,sup}$ and $\texttt{INV}_{sup}\land Q_{m,sup}$ in the superclass;
and $\texttt{INV}_{sub}\land P_{m,sub}$ and $\texttt{INV}_{sub}\land Q_{m,sub}$ in the subclass.
We can conclude that the subclass can be used in any place the superclass is used if the following conditions holds.
\begin{enumerate}
\item \label{LSV-RULE}For each method $m$ of the superclass, it holds that $$(\texttt{INV}_{sub}\land P_{m,sup}\Rightarrow P_{m,sub})\land (\texttt{INV}_{sub}\land Q_{m,sub}\Rightarrow Q_{m,sup})$$
\item \label{PRIVATE-MEM}The objects of both the superclass and the subclass do not share their private memories with other objects.
\item \label{INV-SYMBOL}The definition of the predicate $\texttt{INV}$ is not used during the verification of the client code.
\end{enumerate}
Let $B$ be a superclass, and $D$ be a subclass of $B$. Because of the condition \ref{PRIVATE-MEM}, the invariant of an object always holds
when a method of the object is invoked. From the condition \ref{LSV-RULE} and \ref{INV-SYMBOL}, at any place the precondition of $B\mbox{::}m$ holds,
the precondition of $D\mbox{::}m$ also holds. So $B$ can be substituted with $D$ in any client code of $B$.
This is in fact a variant of the Liskov Substitution Principle.

Now we show that these conditions are a special case of the principle presented in this paper to some extent: for an arbitrary piece of
client code using the superclass, we can
construct an interface implemented by both the superclass and the subclass.
\begin{itemize}
\item   All the function symbols (including \texttt{INV}) defined in the superclass are declared in the interface.
        All the properties about these symbols used in the verification of the client code are expressed as constraint templates.
\item For each method $m$, two function (predicate) symbols are declared: $\texttt{PreSym}_m$ and $\texttt{PostSym}_m$. The specification template of $m$ in the interface is
        $$\{\texttt{INV}\land \texttt{PreSym}_m\}\_\{\texttt{INV}\land \texttt{PostSym}_m\}$$
    There are two constraint templates about these two symbols.
        $$\begin{array}{c}
            \texttt{INV}\land P_{m,sup}\Rightarrow \texttt{PreSym}_m\\
            \texttt{INV}\land \texttt{PostSym}_m\Rightarrow Q_{m,sup}
        \end{array}$$
\end{itemize}
Because this interface contains all the properties about the symbols, the verification of the client code can be done based on this interface specification.
It can be checked that both the superclass and the subclass implement this interface:
\begin{itemize}
\item The superclass defines $\texttt{PreSym}_m$ and $\texttt{PostSym}_m$ as $P_{m,sup}$ and $Q_{m,sup}$ respectively, while
the subclass defines these two function symbols as $P_{m,sub}$ and $Q_{m,sub}$ respectively.
From the conditions \ref{LSV-RULE}, the constrains about  $\texttt{PreSym}_m$ and $\texttt{PostSym}_m$ are satisfied by both the superclass and subclass.
\item The superclass and the subclass give the same definitions to all other function symbols. So other constraints are also satisfied.
\end{itemize}

\begin{figure}
\begin{center}
\begin{boxedminipage}{0.5\textwidth}
{\scriptsize
\begin{tabbing}
\ \ \ \ \=\ \ \ \ \ \=\ \ \ \ \ \ \ \ \=\ \ \ \ \=\\
\textbf{class} SetByListWSize\\
\{\\
\textbf{var:} Node * \texttt{head};\\
\textbf{funcs:}\\
    \>\textbf{attrib} \textbf{SetOf}(\textbf{int}) \texttt{theSet}()  $\triangleq$\\
    \>  \>$\lambda x.x\fldacc D [\texttt{NodeSet}(\textrm{head})]$;\\
    \>\textbf{attrib} \textbf{bool} $\texttt{INV}()\triangleq \texttt{isSList}(\textrm{head})$;\\
    \>\textbf{attrib} \textbf{SetOf}(\textbf{Ptr}) $\texttt{pmem}()\triangleq$\\
    \>  \>$\{\&\textbf{this}\fldacc \textrm{head}\}\cup$\\
    \>  \>$(\lambda x.(\&x\fldacc D)[\texttt{NodeSet}(\textrm{head})])\cup$\\
    \>  \>$(\lambda x.(\&x\fldacc link)[\texttt{NodeSet}(\textrm{head})])$\\
\textbf{method}:\\
    \>SetByList()\\
    \>  \>\textbf{pre} $\rho\land \texttt{INV}()$ \\
    \>  \>\textbf{post} $\rho \land \texttt{INV}() \land (\texttt{theSet}()=\emptyset)$\\
    \>  \>\{head = \textbf{nil};\};\\
    \>\textbf{bool}  isIn(\textbf{int} x)\\
    \>  \>\textbf{pre} $\rho\land \texttt{INV}()$\\
    \>  \>\textbf{post} $\rho \land \texttt{INV}()$\\
    \>  \>  \>$\land(\textbf{ret} = (x\in \textbf{this}\fldacc \texttt{theSet}())$\\
    \>\{\>Node *cur;\\
    \>  \>  cur = head;\\
    \>  \>  while(cur $\neq$ nil $\land$ cur\fldacc D $\neq$ x)\\
    \>  \>  \>cur := cur\fldacc link;\\
    \>  \>return (cur = nil)\\
    \>\};\\
    \>\textbf{int}  getSizeOf()\\
    \>  \>\textbf{pre} $\rho\land \texttt{INV}()$\\
    \>  \>\textbf{post} $\rho \land \texttt{INV}()$\\
    \>  \>  \>$\land(\textbf{ret} = \textbf{sizeOf}(\textbf{this}\fldacc \texttt{theSet}()))$\\
    \>\{\>Node *cur; int l;\\
    \>  \>cur = first; l = 0;\\
    \>  \>while (cur$\neq$ nil)\\
    \>  \>\{cur = cur\fldacc link; l = l + 1;\}\\
    \>  \>return l;\\
    \>\};\\
    \>\textbf{void} Add(int x)\\
    \>  \>\textbf{pre} $\rho \land (\mathfrak{M}(\rho)\cap \texttt{pmem}() = \emptyset)$\\
    \>  \>  \>$\land \texttt{INV}()\land (x\not\in \textbf{this}\fldacc \texttt{theSet}())$\\
    \>  \>\textbf{post} $\rho \land \texttt{INV}()$\\
    \>  \>  \>$\land (\texttt{theSet}()=\overleftarrow{\texttt{theSet}()}\cup\{x\})$\\
    \>\{\>Node *tmp;\\
    \>  \>tmp := alloc(Node); \\
    \>  \>tmp\fldacc D = x; tmp\fldacc link = head;\\
    \>\};\\
\}
\end{tabbing}
}
\end{boxedminipage}
\begin{boxedminipage}{0.5\textwidth}
{\scriptsize
\begin{tabbing}
\ \ \ \ \=\ \ \ \ \ \=\ \ \ \ \ \ \ \ \=\ \ \ \ \=\\
\textbf{class} SetByListWSize \textbf{inherit} \texttt{SetByList}\\
\{\\
\textbf{var:} Node * \texttt{head};  int size;\\
\textbf{funcs:}\\
    \>\textbf{attrib} \textbf{SetOf}(\textbf{int}) \texttt{theSet}()  $\triangleq$\\
    \>  \>$\lambda x.x\fldacc D [\texttt{NodeSet}(\textrm{head})]$;\\
    \>\textbf{attrib} \textbf{bool} $\texttt{INV}()\triangleq\texttt{isSList}(\textrm{head})\land$\\
    \>  \>$\textrm{size}=\textbf{sizeOf}(\texttt{theSet}())$;\\
    \>\textbf{attrib} \textbf{SetOf}(\textbf{Ptr}) $\texttt{pmem}()\triangleq$\\
    \>  \>$\{\&\textbf{this}\fldacc \textrm{head},\&\textbf{this}\fldacc \textrm{size}\}\cup$\\
    \>  \>$(\lambda x.(\&x\fldacc D)[\texttt{NodeSet}(\textrm{head})])\cup$\\
    \>  \>$(\lambda x.(\&x\fldacc link)[\texttt{NodeSet}(\textrm{head})])$\\
\textbf{method}:\\
    \>SetByListWSize()\\
    \>  \>\textbf{pre} $\rho\land \texttt{INV}()$ \\
    \>  \>\textbf{post} $\rho \land \texttt{INV}() \land (\texttt{theSet}()=\emptyset)$\\
    \>  \>\{head = \textbf{nil};\\
    \>  \>\ size = 0;\};\\
    \>\textbf{bool}  isIn(\textbf{int} x)\\
    \>  \>\textbf{pre} $\rho\land \texttt{INV}()$\\
    \>  \>\textbf{post} $\rho \land \texttt{INV}()$\\
    \>  \>  \>$\land(\textbf{ret} = (x\in \textbf{this}\fldacc \texttt{theSet}())$\\
    \>\{\>Node *cur;\\
    \>  \>  cur = head;\\
    \>  \>  while(cur $\neq$ nil $\land$ cur\fldacc D $\neq$ x)\\
    \>  \>  \>cur := cur\fldacc link;\\
    \>  \>return (cur = nil)\\
    \>\};\\
    \>\textbf{int}  getSizeOf()\\
    \>  \>\textbf{pre} $\rho\land \texttt{INV}()$\\
    \>  \>\textbf{post} $\rho \land \texttt{INV}()$\\
    \>  \>  \>$\land(\textbf{ret} = \textbf{sizeOf}(\textbf{this}\fldacc \texttt{theSet}()))$\\
    \>\{\\
    \>  \>\textbf{return} size;\\
    \>\};\\
    \>\textbf{void} Add(int x)\\
    \>  \>\textbf{pre} $\rho \land (\mathfrak{M}(\rho)\cap \texttt{pmem}() = \emptyset)$\\
    \>  \>  \>$\land \texttt{INV}()\land (x\not\in \textbf{this}\fldacc \texttt{theSet}())$\\
    \>  \>\textbf{post} $\rho \land \texttt{INV}()$\\
    \>  \>  \>$\land (\texttt{theSet}()=\overleftarrow{\texttt{theSet}()}\cup\{x\})$\\
    \>\{\>Node *tmp;\\
    \>  \>tmp := alloc(Node); \\
    \>  \>tmp\fldacc D = x; tmp\fldacc link = head;\\
    \>  \>size = size + 1;\\
    \>\};\\
\}
\end{tabbing}
}
\end{boxedminipage}

\end{center}
\caption{The class SetByList and its subclass SetByListWSize}\label{FIG-INHERITENCE}
\end{figure}

\begin{example}
Suppose that there is a class \texttt{SetBySList} as depicted in the left part of Fig.~\ref{FIG-INHERITENCE}, which records all the elements of a finite integer set in a singly linked list.
The invariant \texttt{INV}() says that the member variable $head$ refers to the head node of a singly-linked list.
The \texttt{SetBySList} objects do not share their private memories with other objects.

To accelerate the queries on the size of the set,  we can replace the class \texttt{SetBySList} by a subclass \texttt{SetBySListWSize}, depicted in the right part of Fig.~\ref{FIG-INHERITENCE}.
This subclass has an extra member variables \texttt{size}.
The invariant \texttt{INV}() says that the value of \texttt{size} is just the size of the set.
This property is maintained by all the methods that modify the set.
The specifications of the corresponding methods are literally identical, but the definitions of the symbols
are different. The bodies of these methods are modified w.r.t. the new definitions.
It can be checked that the relation between the subclass \texttt{SetBySListWSize} and the superclass \texttt{SetBySList} satisfies the
conditions given in this subsubsection. So the subclass \texttt{SetBySListWSize} can be used in anyplace the superclass \texttt{SetBySList} is used.

If we found the extra cost of maintaining the value of \texttt{size} is higher than that saved by directly returning \texttt{size} in \texttt{getSize}() of \texttt{SetBySListWSize},
we can get a subclass \texttt{SetBySListWoSize} of \texttt{SetBySListWSize}.
This subclass does not use the member variable \texttt{size} at all, and the methods are same as the ones
in the class \texttt{SetBySList}. In this case, the post-conditions of the methods of \texttt{SetBySListWoSize} is weaker than those of \texttt{SetBySListWSize}.
The principle in this paper shows that \texttt{SetBySListWoSize} is still a subtype of \texttt{SetBySListWSize}.
\hfill$\Box$
\end{example}

\section{Related Works and Conclusions}\label{SEC-CONCLUSION}
`Programming to interfaces' is an important programming paradigm in Object-Oriented programming community.
This paradigm is mainly supported by the runtime-binding facility through class-inheritance or interface-implementation
in OO programming languages. An invocation to a method of an interface (or superclass) may be dynamically bound to a method defined in an implementing class (or a subclass).
The main challenge to specify `programming to interface' code is to deal with the polymorphism
caused by the runtime-binding facility.

Many research works have been proposed to deal with the polymorphism cased by method inheritance and overriding.
Most of the works use the LSP (Liskov Substitution Principle) subtyping rule \cite{Liskov:1994:BNS:197320.197383} to avoid re-verification of the client code.
Once a method has committed to a pre-conditoin/post-condition contract, any redefinition of
this method through overriding must preserve to this commitment.

In \cite{DBLP:conf/esop/Poetzsch-HeffterM99}, the Virtual Method approach is presented to handle method overriding and dynamic dispatch.
Virtual methods represent the common properties of all corresponding subtype methods. All the implementations of a virtual method
must commit to these abstract properties defined in the superclass. However, these abstract properties may be too weak to
verify the client code using the superclass, especially when the superclass is abstract. In \cite{DBLP:conf/popl/ChinDNQ08}, each
method is associated with two specifications: a static specification, which is applied when the dynamic class of the receiver is known,
and a dynamic one, which is used for method invocation with dynamic dispatch. A dynamic specification must be a specification supertype of its static
counterpart, and be a specification supertype of the dynamic specification of each overriding method in its subclasses. A similar
approach is presented in \cite{Parkinson2008}. In \cite{DBLP:journals/fac/SmansJPS10}, a set of axioms are used to specify that the overriding methods must commit to
the specifications of the overridden ones. All the above approaches deal with dynamic dispatch based on the LSP sub-typing rule
(or its variants). As we discussed before, such approaches are not suitable for the `programming to interfaces' paradigm, because the interfaces
(or abstract classes) declare no member variable. The interface methods can not be precisely specified without member variable.

In \cite{DBLP:journals/jlp/DovlandJOS10}\cite{DBLP:journals/scp/DovlandJOS11}, a lazy form of behavioral sub-typing is presented.
The behaviors of the overriding methods are not restricted by the specifications of the overridden methods.
The overriding methods are only required to preserve the `part' of the specifications that actually used to verify the
client codes. This approach still requires the specifications of the overridden methods. So it is not powerful enough to deal with the
`programming to interfaces' paradigm.

Abstract predicate families are used in \cite{Parkinson2008} to specify the methods of super-classes and sub-classes.
These predicates can also be re-defined differently by subclasses. However, they are mainly used to deal with the extend fields in subclasses.
In our paper, abstract function symbols are used to capture the relationship between interfaces and classes.

In this paper, we present a flexible and precise approach to specify and verify code written in
 the `programming to interfaces' paradigm. It addresses the following problems.
\begin{itemize}
 \item \textbf{How to specify an interface.} A set of abstract function/predicate symbols, together with a set of constraints on these symbols, are declared in the interface. The methods of the interface can be precisely specified using these function/predicate symbols.
 \item \textbf{How a class implements an interface.} The class can gives its own definitions to the function/predicate symbols in the interface, as long as the constraints declared in the interface are satisfied.
     The class can give its own implementations to the methods declared in the interface, as long as the method specifications are satisfied w.r.t. the symbol definitions in the class.
\item \textbf{How to verify the client code using interfaces.} The term $\textbf{classOf}(e)$ is used to denote the runtime class of $e$. Given an expression $e$, $\textbf{classOf}(e)\mbox{::}f$ refers to the function definition of $f$ in the class $\textbf{classOf}(e)$. Logical deduction can be performed based on the constraint templates declared in the interfaces, without referring to the classes implementing these interfaces.
\end{itemize}
The class-interface-implementation relation proposed in this paper is flexible. A superclass is viewed as an interface plus an implementation to this interface.
From this view, if a subclass is a subtype of the superclass according the LSP subtyping rule, we can add some function symbols to the interface of the superclass such that
the subclass implements the interface according to the class-interface-implementation relation proposed in this paper. To some extent, the LSP subtyping rule is a special case of our class-interface relation.

While the interface method specifications are abstract, they are  precise enough to deal with some popular real-world examples (See Appendix). The client code can be specified and verified without referring to the implementing classes. However, when more information about the dynamic classes of expressions are known, the client code specifications can be specialized to more precise ones without re-verification. In the `programming to interfaces' paradigm, programmers can give different implementations to an interface to make the client code fulfill different functional features. The approach presented in this paper supports this advantage in that different specifications of a piece of client code can be inferred from the general specification of the client code and the different implementations of the interface.

We also give several real-world examples in the Appendixes to demonstrate the power of our approach.
\bibliographystyle{plain}
\bibliography{oobib}

\newpage
\appendix

\section{The specification of the interface Comparator}
In this section, we given the specification of the interface \texttt{ComparatorOfPoints}, which can be used to compare two points.
The interface \texttt{ComparatorOfPoints} declared in Fig.~\ref{FIG-COMPARATOR}  is similar to the interface
\texttt{java.util.Comparator}. The interface \texttt{java.util.Comparator} is a template and has a type parameter $T$.
As templates are not supported by the small language in this paper,  the parameter $T$ is fixed as \texttt{Point}, so we have \texttt{ComparatorOfPoints}.
Two implementations of this interface are also given, together with a piece client code using this interface.

\subsection{The interface specification}
The specification of the interface \texttt{ComparatorOfPoint} is given in Fig.~\ref{FIG-IMPLEMENTATION}.
It declares an object-function symbol \texttt{LE}. The first three constraint templates declared in the interface say that
\texttt{LE} induces a total order. The fourth constraint template says that
the memory scope of \texttt{LE} is a subset of the private memories of \textbf{this} and the two points being compared.
The method \texttt{Compare} compares two points w.r.t. the order induced by \texttt{LE}.

\begin{figure}
\begin{center}
\begin{boxedminipage}{0.8\textwidth}
\scriptsize
\begin{tabbing}
\ \ \ \ \=\ \ \ \ \=\ \ \ \ \=\\
\textbf{interface} ComparatorOfPoints\\
\{\\
\textbf{funcs:}\\
    \>\textbf{bool} \texttt{LE}(\texttt{Point} v1, \texttt{Point} v2);\\
\textbf{cons:}\\
    \>$\forall v:\texttt{Point}.(v\neq \textbf{nil}\Rightarrow \texttt{LE}(v,v))$;\\
    \>$\forall v_1,v_2:\texttt{Point}.((v_1\neq \textbf{nil}\land v_2\neq \textbf{nil})\Rightarrow (\texttt{LE}(v_1,v_2)\lor \texttt{LE}(v_2,v_1)))$;\\
    \>$\forall v_1,v_2,v_3:\texttt{Point}.(\texttt{LE}(v_1,v_2)\land \texttt{LE}(v_2,v_3)\Rightarrow \texttt{LE}(v_1,v_3))$;\\
    \>$\forall v_1,v_2:\texttt{Point}.((v_1\neq \textbf{nil}\land v_2\neq \textbf{nil})\Rightarrow$\\
    \>  \>$(\mathfrak{M}(\texttt{LE})(v_1,v_2)\subseteq(\textbf{this}\rightarrow \textbf{pmem}()\cup v_1\rightarrow \textbf{pmem}()\cup v_2\rightarrow \textbf{pmem}()))$;\\
\textbf{methods}:\\
    \>\textbf{int} Compare(\texttt{Point} *$o_1$, \texttt{Point} *$o_2$);\\
    \>  \>\textbf{Pre}\ \ $\{\rho\land o_1\neq \textbf{nil}\land o_2\neq \textbf{nil}\}$\\
    \>  \>\textbf{Post}  $\{\rho\land (  (\textbf{ret}\le 0\Rightarrow \texttt{LE}(o_1,o_2))\land(\textbf{ret}\ge 0\Rightarrow \texttt{LE}(o_2,o_1)))\}$\\
\}\\
\end{tabbing}
\end{boxedminipage}
\end{center}
\caption{The specification of the interface ComparatorOfPoints}\label{FIG-COMPARATOR}
\end{figure}

\subsection{Two implementations of \texttt{ComparatorOfPoints}}
In this subsection, we will give two different implementations of the interface \texttt{ComparatorOfPoints}.
The first implementation \texttt{disComp} compares two \texttt{Point} objects by their distances from a given point,
and the second implementation \texttt{XYComp} first compares two \texttt{Point} objects by
their x-coordinates, and then compares their y-coordinates if two points have the same x-coordinate.
These two implementations are depicted in Fig.~\ref{FIG-COMPARATOR-IMP}.
We can check that these two classes implement the interface \texttt{ComparatorOfPoints}.

\begin{figure}
\begin{center}
\begin{boxedminipage}{0.8\textwidth}
\scriptsize
\begin{tabbing}
\ \ \ \ \=\ \ \ \ \=\ \ \ \ \=\ \ \ \ \=\\
\textbf{class} disComp \textbf{impl} ComparatorOfPoints\\
\{\\
\textbf{var:}\\
    \>\textbf{int} x;\\
    \>\textbf{int} y;\\
\textbf{funcs:}\\
    \>\textbf{int} \texttt{FldX}()$\triangleq$ x;\\
    \>\textbf{int} \texttt{FldY}()$\triangleq$ y;\\
    \>$\textbf{bool}\ \texttt{LE}(\texttt{Point}\ o_1, \texttt{Point} \ o_2) \triangleq$\\
    \>  \>$(o_1\fldacc \texttt{FldX}()-x)^2+ (o_1\fldacc \texttt{FldY}()-y)^2 \le $\\
    \>  \>$(o_2\fldacc \texttt{FldX}()-x)^2 + (o_2\fldacc \texttt{FldY}()-y)^2$\\
\textbf{methods}:\\
    \> disComp(int ox; int oy;)\\
    \>  \>\textbf{Pre} \>\>$\rho$\\
    \>  \>\textbf{Post}\>\>$\rho\land(\mathfrak{M}(\rho)\cap \textbf{this}\fldacc \texttt{BLOCK}()=\emptyset)\land$\\
    \>  \>  \>  \>$ox=\texttt{FldX}()\land oy=\texttt{FldY}()$\\
    \>  \>\{ x:= ox; y:=oy; \}\\
    \>\textbf{int} Compare(\texttt{Point} $o_1$, \texttt{Point} $o_2$);\\
    \>\{\>$\textbf{int}\ x_1; \textbf{int}\ y_1; \textbf{int}\ x_2; \textbf{int}\ y_2;$\\
    \>  \>$x_1 = o_1\fldacc \texttt{getX}(); y_1 = o_1\fldacc \texttt{getY}();$\\
    \>  \>$x_2 = o_2\fldacc \texttt{getX}(); y_2 = o_2\fldacc \texttt{getY}();$\\
    \>  \>\textbf{return} $(x_1-x)^2+(y_1-y)^2$\\
    \>  \>  \>  \>\ \ $-(x_2-x)^2-(y_2-y)^2$\\
    \>\}\\
\}
\end{tabbing}
\end{boxedminipage}
\begin{boxedminipage}{0.8\textwidth}
\scriptsize
\begin{tabbing}
\ \ \ \ \=\ \ \ \ \=\ \ \ \ \=\ \ \ \ \=\\
\textbf{class} XYComp \textbf{impl} ComparatorOfPoints\\
\{\\
\textbf{funcs:}\\
    \>$\textbf{bool}\ \texttt{LE}(\textbf{Ptr}(\texttt{Point})\ o_1, \textbf{Ptr}(\texttt{Point})\ o_2) \triangleq$\\
    \>  \>$(o_1\fldacc \texttt{FldX}()< o_2\fldacc \texttt{FldX}()) \lor $\\
    \>  \>$(o_1\fldacc \texttt{FldX}()= o_2\fldacc \texttt{FldX}()\land$\\
    \>  \>$\ \ \ \ \ \ \ o_1\fldacc \texttt{FldY}()\le o_2\fldacc \texttt{FldY}() )$\\
\textbf{methods}:\\
    \> XYComp( )\\
    \>  \>\textbf{Pre} \>\>$\rho$\\
    \>  \>\textbf{Post}\>\>$\rho\land(\mathfrak{M}(\rho)\cap \textbf{this}\fldacc \texttt{BLOCK}()=\emptyset)$\\
    \>  \>\{ \ \  \}\\
    \>\textbf{int} Compare(\texttt{Point} *$o_1$, \texttt{Point} *$o_2$);\\
    \>\{\\
    \>  \>$\textbf{int}\ x_1; \textbf{int}\ y_1;$\\
    \>  \>$\textbf{int}\ x_2; \textbf{int}\ y_2;$\\
    \>  \>$x_1 = o_1\fldacc \texttt{getX}(); y_1 = o_1\fldacc \texttt{getY}();$\\
    \>  \>$x_2 = o_2\fldacc \texttt{getX}(); y_2 = o_2\fldacc \texttt{getY}();$\\
    \>  \>\textbf{return} $(x_1<x_2)\lor$\\
    \>  \>  \>  \>\ \ $((x_1=x_2)\land(y_1\le y_2))$\\
    \>\}\\
\}\\
\end{tabbing}
\end{boxedminipage}
\end{center}
\caption{Tow implementations of the interface ComparatorOfPoints}\label{FIG-COMPARATOR-IMP}
\end{figure}

\begin{figure}
\begin{boxedminipage}{0.5\textwidth}
{\scriptsize
\begin{tabbing}
\mbox{}\ \ \ \ \ \=\ \ \ \ \ \=\ \ \ \ \ \=\ \ \ \ \ \=\ \ \ \ \ \=\\
class ArraySort\{\\
\textbf{var:}\\
    \>\texttt{Point} \verb"arr"[10];\\
\textbf{funcs:}\\
    \>$\textbf{SetOf}(\textbf{Ptr})\ \texttt{pmem}()\triangleq \lambda x.(\&\textbf{this}\fldacc \verb"arr"[x])[0..9]$;\\
    \>$\texttt{Point}\ \texttt{get}(\textbf{int}\ i)\triangleq \verb"arr"[i]$;\\
\textbf{method:}\\
    \>$\dots\ \ \dots\ \ \dots\ \ $\\
    \>\textbf{void} Sort(\texttt{ComparatorOfPoints} cmp)\\
    \>\textbf{Pre}\ \   $\rho\land(\forall x\in 0..9 (\texttt{get}(x)\neq \textbf{nil}))\land \texttt{MemoryLayout}$\\
    \>\textbf{Post}  $\rho\land \forall i\in[0..9](\texttt{cmp}\fldacc \texttt{LE}(\texttt{get}(i),\texttt{get}(i+1)))$\\
    \>\{\>\textbf{int} i,j,r; \texttt{Point} tmp;\\
    \>  \>i = 9;\\
    \>  \>\textbf{while} ( i $>$ 0 )\{\\
    \>  \>    \>j = 0;\\
    \>  \>    \>\textbf{while} ( j $<$ i-1 )\\
    \>  \>    \>\{\> r = cmp$\fldacc \texttt{Compare}(\verb"arr"[j],\verb"arr"[j+1])$;\\
    \>  \>    \>  \>\textbf{if}( r $>$ 0 )\{\\
    \>  \>    \>  \>  \>tmp = \verb"arr"[j];\\
    \>  \>    \>  \>  \>\verb"arr"[j]=\verb"arr"[j+1];\\
    \>  \>    \>  \>  \>\verb"arr"[j+1]=tmp;\\
    \>  \>    \>  \>\}\\
    \>  \>    \>  \>\textbf{else}\\
    \>  \>    \>  \>  \>\textbf{skip};\\
    \>  \>    \>  \>j = j+1;\\
    \>  \>    \>\}\\
    \>  \>\}\\
    \>  \>i=i-1;\\
    \>  \}\\
\}\\
\end{tabbing}
}
\end{boxedminipage}
\begin{minipage}{0.4\textwidth}
{\scriptsize
$\textsf{MemoryLayout}$ is the abbreviation for
$$\begin{array}{l}
(\mathfrak{M}(\rho)\cap \textbf{this}\fldacc \texttt{pmem}()=\emptyset)\land\\
(\texttt{cmp}\fldacc \texttt{pmem}()\cap\textbf{this}\fldacc \texttt{pmem}() =\emptyset)\land\\
\lambda x.(\verb"arr"[x]\fldacc \texttt{pmem}())[0..9]\cap \textbf{this}\fldacc \texttt{pmem}()  =\emptyset
\end{array}$$

The invariant of the outer loop is\\
$$
\begin{array}{l}
\rho\land \texttt{MemoryLayout}\land\\
\forall x\in (i..8).(\texttt{cmp}\fldacc \texttt{LE}(\verb"arr"[x],\verb"arr"[x+1]))
\end{array}
$$
\\
\\
The invariant of the inner loop is\\
$$
\begin{array}{l}
\rho\land \texttt{MemoryLayout}\land\\
\forall x\in (i..8).(\texttt{cmp}\fldacc \texttt{LE}(\verb"arr"[x],\verb"arr"[x+1]))\land\\
\forall x\in (0..j-1).(\texttt{cmp}\fldacc \texttt{LE}(\verb"arr"[x],\verb"arr"[j]))
\end{array}
$$
}
\end{minipage}
\caption{The sorting algorithm using Comparator, and its specification}\label{FIG-COMPARATOR-CONTEXT}
\end{figure}

\subsection{The client code using \texttt{ComparatorOfPoint}}
The class \texttt{ArraySort} depicted in Fig.~\ref{FIG-COMPARATOR-CONTEXT} has a method \texttt{Sort} which compares
the \texttt{Point} objects in the list using the interface \texttt{ComparatorOfPoints}.

The invariants of the two while-statements are depicted on the right part of Fig.~\ref{FIG-COMPARATOR-CONTEXT}. Using these invariants, the function can be verified
based on the properties about $\texttt{LE}$ presented in Fig.~\ref{FIG-COMPARATOR}. We can sort the objects in the array \verb"arr" by invoking the method \texttt{Sort} using different comparator as parameter. If the real parameter of \texttt{Sort} is the object created by the following statement
$$\texttt{cmp} = \textbf{new}\ \texttt{disComp}(5,4);$$
the \texttt{Point} objects in \verb"arr" are sorted according to the distances from the point $(5,4)$. If the real parameter is created by
$$\texttt{cmp} = \textbf{new}\ \texttt{XYComp}( );$$
the \texttt{Point} objects are sorted according to the X-coordinates first, and then the Y-coordinates for objects with the same X-coordinates.

\newpage
\section{The Set interface and its Iterator interface}\label{SEC-ITERATOR}
The Iterator pattern decouples the codes traversing elements in a container from the implementation of the container.
This section shows how to specify an Iterator over a set container.

\begin{figure}
\begin{center}
\begin{boxedminipage}{1\textwidth}
\scriptsize
\begin{tabbing}
\ \ \ \ \=\ \ \ \ \=\ \ \ \ \ \ \ \ \=\\
\textbf{interface} Iterator\\
\{\\
\textbf{funcs:}\\
    \>SetOfInt  theContainer( );\\
    \>\textbf{SetOf}(\textbf{int})  passedSet( );\\
    \>\textbf{bool}  SOUND( );\\
\textbf{cons:}\\
    \>$\texttt{INV}()\land \texttt{SOUND}() \Rightarrow (\texttt{theContainer}() \neq \textbf{nil}\land \texttt{theContainer}()\fldacc \texttt{INV}())$;\\
    \>$\texttt{INV}()\land \texttt{SOUND}() \Rightarrow (\texttt{passedSet}() \subseteq \texttt{theContainer}(){\fldacc}\texttt{theSet}())$;\\
    \>$\texttt{INV}()\land \texttt{SOUND}() \Rightarrow (\mathfrak{M}(\texttt{passedSet})() \subseteq (\texttt{pmem}() \cup \texttt{theContainer}(){\fldacc}\texttt{pmem}()))$;\\
    \>$\texttt{INV}()\land \texttt{SOUND}() \Rightarrow (\mathfrak{M}(\texttt{SOUND})() \subseteq (\texttt{pmem}() \cup \texttt{theContainer}(){\fldacc}\texttt{pmem}()))$;\\
\textbf{methods:}\\
    \>\textbf{bool}    hasNext( );\\
    \>  \>\textbf{pre} \>$\rho \land \texttt{SOUND}()$\\
    \>  \>\textbf{post}\>$\rho\land \texttt{SOUND}() \land \textbf{ret} = (\texttt{passedSet}()=\texttt{theContainer}()\fldacc \texttt{theSet}())$\\
    \>\textbf{int}    Next();\\
    \>  \>\textbf{pre}  \>$\rho \land (\mathfrak{M}(\rho) \cap \texttt{pmem}() = \emptyset) \land \texttt{SOUND}()\land\texttt{passedSet}()\neq \texttt{theContainer}()\fldacc \texttt{theSet}()$\\
    \>  \>\textbf{post}\>$\rho\land \texttt{SOUND}( ) \land (\texttt{passedSet}() = \overleftarrow{\mbox{\texttt{passedSet}()}} \cup \{\textbf{ret}\}) \land\textbf{ret}\not\in \overleftarrow{\mbox{passedSet()}} \land $\\
    \>  \>  \>\ \ $\textbf{ret} \in \texttt{theContainer}(){\fldacc}\texttt{theSet}() \land\texttt{theContainer}() = \overleftarrow{\mbox{\texttt{theContainer}()}}$\\
\}\\
\textbf{interface} SetOfInt\\
\{\\
\textbf{funcs}:\\
    \>\textbf{attribute} \textbf{SetOf}(\textbf{int}) theSet();\\
\textbf{methods}:\\
    \> \>$\dots\ \dots$\\
    \>Iterator createIterator();\\
    \>  \>\textbf{pre} \>$\rho$\\
    \>  \>\textbf{post}\>$\rho\land\textbf{ret}\neq\textbf{nil}\land\textbf{ret}{\fldacc}\texttt{INV}()\land\textbf{ret}{\fldacc}\texttt{theContainer}()=\textbf{this}\land \textbf{ret}\fldacc \texttt{SOUND}()\land$\\
    \>  \>  \>$\textbf{ret}{\fldacc}\texttt{passedSet}() = \emptyset$\\
\}
\end{tabbing}
\end{boxedminipage}
\end{center}
\caption{The interfaces \texttt{SetOfInt} and \texttt{Iterator}}\label{FIG-ITERATOR-SPEC}
\end{figure}

\subsection{The specification of the interfaces}
The interface specifications of \texttt{Iterator} and \texttt{SetOfInt} is given in Fig.~\ref{FIG-ITERATOR-SPEC}.
The interface \texttt{SetOfInt} has an attribute function symbol \texttt{theSet}, which denotes the set represented by the
object. The methods for set operations, e.g. add, delete, are omitted. The method \texttt{createIterator} creates an iterator of \textbf{this} object.
The interface \texttt{Iterator} declares 3 function symbols.
\begin{itemize}
\item \texttt{theContainer}() denotes the set to which this iterator is attached;
\item \texttt{passedSet}() denotes the set of integers in \texttt{theContainer}() that have been traversed by the method \texttt{Next}();
\item \texttt{SOUND}( ) specify that \texttt{theContainer}() refers to a valid object of the set.
\end{itemize}
The method \texttt{hasNext}() test whether all the elements are traversed. The method \texttt{Next}() returns an element that
has not been traversed.

\subsection{The classes implementing the interfaces}
Each class implementing the interface \texttt{SetOfInt} should have a method to create corresponding iterators. An iterator can be used to traverse
all the elements in the set that creates this iterator.
The class \texttt{SingleListSetOfInt} given in Fig.~\ref{FIG-ITERATOR-IMP} implements the interface \texttt{SetOfInt}. Its corresponding iterator class
\texttt{IteratorOfSLS}, which is an implementation of the interface \texttt{Iterator},  is also given in Fig.~\ref{FIG-ITERATOR-IMP}.

\begin{figure}
\begin{center}
\begin{boxedminipage}{1.0\textwidth}
{\scriptsize
\begin{tabbing}
\ \ \ \ \=\ \ \ \ \ \=\ \ \ \ \ \ \ \ \=\ \ \ \ \=\\
\textbf{class} SetOfIntSL \textbf{impl} SetOfInt\\
\{\\
\textbf{var:}\\
    \>Node *head;\\
\textbf{funcs:}\\
    \>\textbf{attribute} \textbf{SetOf}(\textbf{int}) theSet() $\triangleq \lambda x.x{\fldacc}\texttt{data}[\texttt{NodeSet}(\texttt{head}{\fldacc}\texttt{link})]$;\\
    \>\textbf{bool}  INV() $\triangleq$ IsSList(head) $\land$ head $\neq$ \textbf{nil};\\
    \>\textbf{SetOf}(\textbf{Ptr}) pmem() $\triangleq \lambda x.\&(x{\fldacc}\texttt{link})[\texttt{NodeSet}(\texttt{head})]$;\\
    \>Node * getHead() $\triangleq \texttt{head}$;\\
\textbf{method}:\\
    \>  \>  \>$\dots\ \ \dots$\\
    \>\textbf{int} createIterator( )\\
    \>\{ \>Iterator it = new IteratorOfSLS(\textbf{this}, head);\ \ \ \textbf{return} it; \}\\
\}\\
\\
\textbf{class} IteratorOfSLS \textbf{impl} Iterator\\
\{\\
\textbf{var:}\\
    \>SetOfIntSL thecontainer;\\
    \>Node *cur;\\
\textbf{funcs:}\\
    \>$\textbf{bool } \texttt{SOUND}() \triangleq \texttt{thecontainer}\neq \textbf{true} \land \texttt{thecontainer}\fldacc \texttt{INV}()$\\
    \>SetOfInt  theContainer( ) $\triangleq \texttt{thecontainer};$\\
    \>\textbf{SetOf}(\textbf{int})\ \ passedSet( ) $\triangleq \texttt{theContainer}(){\fldacc}\texttt{theSet}() - \lambda x.x\fldacc \texttt{data}[\texttt{NodeSet}(\texttt{cur}{\fldacc}\texttt{link})]$;\\
\textbf{method}:\\
    \>InteratorOfSLS(SetOfIntSL *container, Node *head)\\
\>    \>\textbf{pre} \>$\rho$ $\land$ ($\mathfrak{M}$($\rho$) $\cap$ pmem() = $\emptyset$) $\land$ container{\fldacc}INV() $\land$ head = container{\fldacc}getHead()\\
\>    \>\textbf{post}\>$\rho$ $\land$ SOUND() $\land$ theContainer() = container $\land$ passedSet() = $\emptyset$\\
\>    \>\{\\
\>    \>  \>thecontainer = container; cur = head;\\
\>    \>\}\\
    \>\textbf{bool}    hasNext( ) \{\textbf{return} (cur{\fldacc}link $\neq$ nil);\}\\
    \>\textbf{int}    Next() \{\ \ cur = cur{\fldacc}link; \textbf{return} cur{\fldacc}data;\}\\
\}\\
\end{tabbing}
}
\end{boxedminipage}
\end{center}
\caption{The classes implementing SetOfInt and Iterator}\label{FIG-ITERATOR-IMP}
\end{figure}

\subsection{The client code using the interfaces}
A piece of client code using the interface \texttt{Iterator} is given in Fig.~\ref{FIG-ITERATOR-CONTEXT}. The
invariant of the while statement is
$$\begin{array}{l}
\texttt{set} = \texttt{it}\fldacc \texttt{theContainer}() \land \texttt{it}\fldacc \texttt{SOUND}()\\
\texttt{bNotFinished} = (\texttt{it}\fldacc \texttt{passedNet}()\subset \texttt{set}\fldacc \texttt{theSet}())\land\\
\texttt{bHasNegative} = \neg(\forall x\in \texttt{it}\fldacc \texttt{passedSet}().(x\ge0))
\end{array}$$
It can be proved that the post-condition of this code segment is
$$\texttt{bHasNegative} == \neg(\forall x\in \texttt{set}\fldacc \texttt{theSet}().(x\ge0))$$
\begin{figure}
\begin{center}
\begin{boxedminipage}{1.0\textwidth}
{\scriptsize
\begin{tabbing}
\ \ \ \ \=\ \ \ \ \ \=\ \ \ \ \ \ \ \ \=\ \ \ \ \=\\
SetOfInt set;\ \ \ Iterator it;\ \ \ \textbf{bool}    bHasNegative;\ \ \ \textbf{bool} bNotFinished;\\
\\
set = new SetOfIntSL();\\
set{\fldacc}add(1);\\
$\dots$\ \ \ $\dots$\ \ $\dots$\ \ \ $\dots$\\
it = set{\fldacc}createIterator();\\

bHasNegative = false;\\
bFinished = it{\fldacc}hasNext();\\
\textbf{while} (bNotFinished)\\
\{\\
    \>k = it{\fldacc}Next();\\
    \>\textbf{if} (k $<$ 0)\\
    \>  \>bHasNegative = true;\\
    \>\textbf{else}\\
    \>  \>\textbf{skip};\\
    \>bNotFinished = it{\fldacc}hasNext();\\
\}\\
\\
\end{tabbing}
}
\end{boxedminipage}
\end{center}
\caption{The context using the \texttt{Iterator} interface and the \texttt{SetOfInt} interface}\label{FIG-ITERATOR-CONTEXT}
\end{figure}

\newpage
\section{The observer pattern}
In this section, we will present the specifications and implementations of a simplified version of the class \texttt{Observable} and the interface \texttt{Observer}
in the Java package \texttt{java.util}. Roughly speaking, \texttt{Observable} and \texttt{Observer} work as follow.
An \texttt{Observable} object is attached with a set of \texttt{Observer} objects. If the \texttt{Observable} object is modified,
the method \texttt{notifyObservers} is invoked to notify all the \texttt{Observer} objects. For each \texttt{Observer} object, the method \texttt{updateData}() is invoked
such that the \texttt{Observer} object is accordingly modified to maintain a consistence relation with the \texttt{Observable} object.
In this design pattern, the implementations of \texttt{Observer} are dependent on the implementation of \texttt{Observable}. That is,
an implementation of \texttt{Observer} can only observe a specific implementation of \texttt{Observable}.

In the package \texttt{java.util}, \texttt{Observable} is a class and some basic methods are already implemented.
People should implement a subclass of \texttt{Observable} to fulfill some expected functional features.
Because inheritance is not supported in the small language, \texttt{Observable} is declared as an interface and the class implementing \texttt{Observable}
should implement the basic methods by itself.
This section can also be viewed as an example of dealing with subclassing and inheritance: we can
specify the interface of a superclass, and make the subclasses implement the interface.

\subsection{The specifications  of the interfaces \texttt{Observable} and \texttt{Observer}}
The interfaces \texttt{Observable} and \texttt{Observer} are specified in Fig.~\ref{FIG-OBSERVER-SPEC}.
The interface \texttt{Observable} declares two function symbols.
\begin{itemize}
\item the attribute function symbol \texttt{theObservers} yields the set of observers.
\item the object-function symbol \texttt{bIndObservers}() specify that the private memories of the observers are disjoint with each other.
\end{itemize}
Two methods are declared in the interface: \texttt{addObserver} and \texttt{notifyObservers}. The first one adds a new observer into
the observer set; the second one is to be invoked when the \texttt{Observable} object is modified, and all the observers must update their
internal data to keep consistent with the \texttt{Observable} object.

The interface \texttt{Observer} declares two function symbols. The function symbol \texttt{bCanObserve} tests whether this observer
can observe the subject; the function symbol \texttt{bConsistantWith} specifies that the observer keeps consistent with the subject being observed.
The method \texttt{Update}() modifies the internal data of \textbf{this} observer to keep consistent with the real parameter.

\begin{figure}
\begin{center}
\begin{boxedminipage}{1\textwidth}
\scriptsize
\begin{tabbing}
\ \ \ \ \=\ \ \ \ \=\ \ \ \ \ \ \ \ \=\\
\textbf{interface} Observable\\
\{\\
\textbf{funcs:}\\
    \>\textbf{attribute} \textbf{SetOf}(\texttt{Observer})  \texttt{theObservers}();\\
    \>\textbf{bool} bIndObservers();\\
\textbf{cons:}\\
    \>    \>$\texttt{bIndObservers}()\Rightarrow \forall x,y\in \texttt{theObservers}().(x\neq y\Rightarrow x\fldacc \texttt{pmem}()\cap y\fldacc \texttt{pmem}()=\emptyset)$;\\
\textbf{methods}:\\
    \>  \>  \>$\dots\ \ \dots$\\
    \>\textbf{void}    addObserver(Observer ob);\\
    \>  \>\textbf{pre} \>\{$\rho\land (\mathfrak{M}(\rho)\cap \mathfrak{M}(\texttt{theObservers})()=\emptyset)\land \texttt{ob}\fldacc \texttt{bCanObserve}(\textbf{this})\land$\\
    \>  \>              \>\ \ $\texttt{bIndObservers}()\land \forall x\in \texttt{theObservers}().(o\fldacc \texttt{pmem}()\cap x\fldacc \texttt{pmem}()=\emptyset) $\}\\
    \>  \>\textbf{post}\>\{$\rho\land ({\texttt{theObservers}()}=\overleftarrow{\texttt{theObservers}()}\cup \{\texttt{ob}\})\land \texttt{bIndObservers}()$\}\\
    \>\textbf{void}    notifyObservers( );\\
    \>  \>\textbf{pre} \>\{$\rho\land \forall x\in \texttt{theObservers}().(\mathfrak{M}(\rho)\cap x\fldacc \texttt{pmem}()=\emptyset)\land \texttt{bIndObservers}()$\}\\
    \>  \>\textbf{post}\>\{$\rho\land \forall x\in \texttt{theObservers}().(x\fldacc \texttt{bConsistantWith}(\textbf{this}))\land \texttt{bIndObservers}()$\}\\
\}\\
\\
\textbf{interface} Observer\\
\{\\
\textbf{funcs}:\\
    \>\textbf{bool} \texttt{bCanObserve}(Observable *sub);\\
    \>\textbf{bool} \texttt{bConsistantWith}(Observable *sub);\\
\textbf{cons:}\\
    \>$\forall x:\texttt{Observable}(\texttt{bCanObserve}(x)\Rightarrow \mathfrak{M}(\texttt{bConsistantWith})(x)\subseteq \texttt{pmem}()\cup x\fldacc \texttt{pmem}())$\\
\textbf{methods}:\\
    \>\textbf{void} updateData(Observable *\texttt{sub});\\
    \>  \>\textbf{pre} \> \{$\rho\land (\mathfrak{M}(\rho)\cap \texttt{pmem}()=\emptyset)\land(\texttt{sub}\neq \textbf{nil})\land \texttt{bCanObserve}(\texttt{sub})$\}\\
    \>  \>\textbf{post}\> \{$\rho\land \texttt{bConsistantWith}(\texttt{sub})\}$\\
\}
\end{tabbing}
\end{boxedminipage}
\end{center}
\caption{The interfaces Observable and Observer}\label{FIG-OBSERVER-SPEC}
\end{figure}

\begin{figure}
\begin{center}
\begin{boxedminipage}{1\textwidth}
\scriptsize
\begin{tabbing}
\ \ \ \ \=\ \ \ \ \=\ \ \ \ \ \ \ \ \=\\
\textbf{class} TextModel \textbf{impl} Observable\\
\{\\
\textbf{var:}\\
    \>String str;\\
    \>\texttt{SetOfObserver}   obSet;\\
\textbf{funcs:}\\
    \>\textbf{attribute} \textbf{SetOf}(\texttt{Observer})  $\texttt{theObservers}()\triangleq \texttt{obSet}\fldacc \texttt{theSet}()$;\\
    \>\textbf{bool} $\texttt{bIndObservers}()\triangleq
                    \forall x,y\in \texttt{theObservers}().(x\neq y\Rightarrow x\fldacc \texttt{pmem}()\cap y\fldacc \texttt{pmem}()=\emptyset)$;\\
    \>String $\texttt{getTextString}()\triangleq \textrm{str}$;\\
\textbf{methods}:\\
    \>TextMode()\\
    \>  \>\textbf{pre}       \>\{$\rho$\}\\
    \>  \>\textbf{post}      \>\{$\rho\land (\textbf{this}\fldacc \texttt{theObservers}() = \emptyset)\land \textbf{this}\fldacc \texttt{getTextString}()=``\ "$\}\\
    \>  \>\{ str = ``\ ";\ \ obSet = \textbf{new} \texttt{SetOfObserverSL}();\ \}\\
    \>void SetText(String s)\\
    \>  \>\textbf{pre}       \>\{$\rho\land (\mathfrak{M}(\rho)\cap \texttt{pmem}()=\emptyset)$\}\\
    \>  \>\textbf{post}      \>\{$\rho\land (\texttt{theObservers}() = \overleftarrow{ \texttt{theObservers}()})\land \texttt{getTextString}()= s$\}\\
    \>  \>\{ str = s; \}\\
    \>  \>  \>$\dots\ \ \dots$\\
    \>\textbf{void}    addObserver(Observer ob);\\
    \>  \>  \{\ $\texttt{obSet}\fldacc \texttt{AddEle}(\texttt{ob})$;\ \}\\
    \>\textbf{void}    notifyObservers( );\\
    \>\{  \>Iterator it; Observer o; bool tmp;\\
    \>  \>it = obSet.createIterator();\\
    \>  \>tmp = it.hasNext();\\
    \>  \>\textbf{while} (tmp)\\
    \>  \>\{\ \ o = it.Next(); o\fldacc updateData(this); tmp = it.hasNext();\ \ \}\\
    \>\}\\
\}\\
\\
\textbf{class} LengthObserver \textbf{impl} Observer\\
\{\\
\textbf{var}:\\
    \>\textbf{int} len;\\
\textbf{funcs}:\\
    \>\textbf{int} $\texttt{length}()\triangleq \texttt{len}$;\\
    \>\textbf{bool} $\texttt{bCanObserve}(\texttt{Observable}\ \texttt{sub})\triangleq (\textbf{classOf}(\texttt{sub}) = \texttt{TextModel})$;\\
    \>\textbf{bool} $\texttt{bConsistantWith}(\texttt{Observable}\ \texttt{sub})\triangleq ( \texttt{length}() = \texttt{strlength}(\texttt{sub}\fldacc \texttt{getTextString}()))$;\\
\textbf{methods}:\\
    \>LengthObserver()\\
    \>  \>pre   \>\{$\rho$\}\\
    \>  \>post  \>\{$\rho\land \texttt{len} = 0;$\}\\
    \>  \>\{ len = 0; \}\\
    \>\textbf{void} updateData(Observable \texttt{sub})\\
    \>  \>\{\\
    \>  \>   \>String s;  TextMode tm;\\
    \>  \>   \>tmp = (TextModel)sub;  s = tm\fldacc getTextString(); len = s{\fldacc}length();\\
    \>  \>\}\\
\}\\
\\
\textbf{class} LineNumObserver \textbf{impl} Observer\\
\{\\
\textbf{var}:\\
    \>\textbf{int} lineNum;\\
\textbf{funcs}:\\
    \>\textbf{int} $\texttt{lNo}()\triangleq \texttt{lineNum}$;\\
   \>\textbf{bool} $\texttt{bCanObserve}(\texttt{Observable}\ \texttt{sub})\triangleq (\textbf{classOf}(\texttt{sub}) = \texttt{TextModel})$;\\
   \>\textbf{bool} $\texttt{bConsistantWith}(\texttt{Observable}\ \texttt{sub})\triangleq ( \texttt{lNo}() = \texttt{NumOfSubString}(\texttt{sub}\fldacc \texttt{getTextString}(),``\backslash\texttt{n}"))$;\\
\textbf{methods}:\\
    \>LineNumObserver()\\
    \>  \>pre   \>\{$\rho$\}\\
    \>  \>post  \>\{$\rho\land \texttt{lineNum} = 0;$\}\\
    \>  \>\{ lineNum = 0; \}\\
    \>\textbf{void} updateData(Observable \texttt{sub});\\
    \>  \>\{\\
    \>  \>  \>String s;  TextMode tm;\\
    \>  \>  \>tmp = (TextModel)sub;  s = tm\fldacc getTextString(); lineNum = s{\fldacc}NumOfSubStr($"\backslash \texttt{n}"$);\\
    \>  \>\}\\
\}

\end{tabbing}
\end{boxedminipage}
\end{center}
\caption{The implementations of \texttt{Observable} and \texttt{Observer}}\label{FIG-OBSERVER-IMP}
\end{figure}

\subsection{The implementations of \texttt{Observable} and \texttt{Observer}}
In this subsection, we give a class \texttt{TextModel} implementing \texttt{Observable}, and two classes, \texttt{LengthObserver}
and \texttt{LineNumObserver}, implementing \texttt{Observer}.
A \texttt{TextModel} object holds a string. The method \texttt{SetText} sets the text string of a \texttt{TextModel} object.
An object of \texttt{LengthObserver} ( or \texttt{LineNumObserver} ) can observe a \texttt{TextModel} object.
It records the number of lines ( or the length of the text string ) of the \texttt{TextModel} object being observed.

A container class \texttt{SetOfObserverSL} and its iterator \texttt{IteratorOfSObLS} is used to record and manipulate the set of observers.
Their specifications and implementations are similar to the ones in Section~\ref{SEC-ITERATOR}. The \texttt{String} class represents
character strings. The method \texttt{length}() returns the length of the string, the method \texttt{NumOfSubStr} returns the number of occurrences of the real parameter
in the string.

\subsection{The client code using the \texttt{Observable} and \texttt{Observer}}
The piece of code depicted in Fig.~\ref{FIG-OBSERVER-CONTEXT} using these two interfaces and their implementations.
It first creates a \texttt{TextModel} object, a \texttt{LengthObserver} object, and a \texttt{LineNumObserver} object.
The later two objects are set to observe the \texttt{TextModel} object.
After some manipulations are performed on the \texttt{TextModel}, the method \texttt{NotifyObservers} is invoked.
After invocation, these two observers are consistent with the \texttt{TextModel} object, i.e. the following property holds.
$$\begin{array}{l}
\texttt{lenOb}\fldacc \texttt{ConsistentWith}(\texttt{tm}) \land \texttt{lnOb}\fldacc \texttt{ConsistentWith}(\texttt{tm})\\
\end{array}$$
Because the runtime classes of \texttt{lenOb} and \texttt{lnOb} are respectively \texttt{LengthObserver} and \texttt{LineNumObserver}, the above
property is equivalent to
$$\begin{array}{l}
(\texttt{lnOb}\fldacc \texttt{lNo}() = \texttt{NumOfSubString}(\texttt{sub}\fldacc \texttt{getTextString}(),``\backslash\texttt{n}))\land\\
(\texttt{lenOb}\fldacc \texttt{length}() = \texttt{strlength}(\texttt{sub}\fldacc \texttt{getTextString}()))
\end{array}$$

\begin{figure}
\begin{center}
\begin{boxedminipage}{1\textwidth}
\scriptsize
\begin{tabbing}
\ \ \ \ \=\ \ \ \ \=\ \ \ \ \ \ \ \ \=\\
Observable tm = new TextModel();\\
LengthObserver lenOb = new LengthObserver();\\
LineNumObserver lnOb = new LineNumObserver();\\

tm{\fldacc}AddObserver(lenOb);\\
tm{\fldacc}AddObserver(lenOb);\\
\\
tm{\fldacc}SetText(``A new string$\backslash$n is set to the$\backslash$n TextModel object'');\\
$\dots,\dots,\dots$\\
tm{\fldacc}NotifyObservers();\\

\end{tabbing}
\end{boxedminipage}
\end{center}
\caption{A piece of code using the interface \texttt{Observer} and \texttt{Observable}}\label{FIG-OBSERVER-CONTEXT}
\end{figure}

\end{document}